%% file: main.tex
\documentclass[sigconf, nonacm]{acmart}
\usepackage{etoolbox}
\usepackage{multirow}
\usepackage{makecell}
\usepackage{xspace}
\usepackage{tabularx}
\usepackage{subcaption}
\usepackage{graphicx}
\usepackage[ruled, noend, linesnumbered]{algorithm2e}
\usepackage{algpseudocode}
\usepackage{xcolor}
\usepackage{hyphenat}
\usepackage{balance}
\AtBeginDocument{%
  }

\newtheorem{example}{Example}

\newcommand{\eg}{\textit{e.g.,}\xspace}
\newcommand{\ie}{\textit{i.e.,}\xspace}
\newcommand{\todo}[1]{}
\newcommand{\micronn}{\nohyphens{\textsc{MicroNN}}\xspace}
\begin{document}

\title{MicroNN: An On-device Disk-resident Updatable Vector Database}

\author{Jeffrey Pound}
\affiliation{%
  \institution{Apple}
  \city{Waterloo}
  \state{ON}
  \country{Canada}
}

\author{Floris Chabert}
\affiliation{%
  \institution{Apple}
  \city{Miami}
  \state{FL}
  \country{USA}
}

\author{Arjun Bhushan}
\affiliation{%
  \institution{Apple}
  \city{Cupertino}
  \state{CA}
  \country{USA}
}

\author{Ankur Goswami}
\affiliation{%
  \institution{Apple}
  \city{Seattle}
  \state{WA}
  \country{USA}
}

\author{Anil Pacaci}
\affiliation{%
  \institution{Apple}
  \city{Seattle}
  \state{WA}
  \country{USA}
}

\author{Shihabur Rahman Chowdhury}
\affiliation{%
  \institution{Apple}
  \city{Seattle}
  \state{WA}
  \country{USA}
}

\renewcommand{\shortauthors}{Pound et al.}

\input{sections/abstract}



\maketitle

\input{sections/introduction}
\input{sections/motivation}

\input{sections/overview}
\input{sections/indexing}
\input{sections/query}
\input{sections/experiments}
\input{sections/conclusion}
\bibliographystyle{ACM-Reference-Format}
\balance
\bibliography{refdb}

\end{document}

%% file: sections/abstract.tex
\begin{abstract}
Nearest neighbour search over dense vector collections has important applications in information retrieval, retrieval augmented generation (RAG), and content ranking. Performing efficient search over large vector collections is a well studied problem with many existing approaches and open source implementations. However, most state-of-the-art systems are generally targeted towards scenarios using large servers with an abundance of memory, static vector collections that are not updatable, and nearest neighbour search in isolation of other search criteria. We present \emph{Micro Nearest Neighbour} (\micronn), an embedded nearest-neighbour vector search engine designed for scalable similarity search in low-resource environments.
\micronn\ addresses the problem of on-device vector search for real-world workloads containing updates and hybrid search queries that combine nearest neighbour search with structured attribute filters. In this scenario, memory is highly constrained and disk-efficient index structures and algorithms are required, as well as support for continuous inserts and deletes. \micronn\ is an embeddable library that can scale to large vector collections with minimal resources. \micronn is used in production and powers a wide range of vector search use-cases on-device. \micronn takes less than 7 ms to retrieve the top-100 nearest neighbours with 90\% recall on publicly available million-scale vector benchmark while using $\approx$10 MB of memory.
\end{abstract}

%% file: sections/introduction.tex
\section{Introduction}\label{sec:introduction}
Advances in modern machine learning models are enabling rich representation of text, images, and other assets as dense vectors that capture semantically meaningful information~\cite{hossain2019comprehensive, minaee2021deep}. These embedding models are trained to produce numerical vectors that are similar (according to a given metric) if the underlying text or images are similar. For example, two pictures of a similar car would both have embedding vectors that are similar. In a joint text and image embedding space, such as those produced by a CLIP model~\cite{pmlr-v139-radford21a}, a picture of a black cat playing with yarn would produce an embedding vector that is similar to the embedding vector produced by a the sentence ``\emph{a black cat playing with yarn}''. Embeddings encode semantics in a latent representation and similarity search over these vectors is becoming an integral part of industrial search engines~\cite{grbovic2018real, haldar2019applying, hashemi2021neural, qin2021mixer, 10.1145/3534678.3539071, 10.1145/3580305.3599782} and recommendations systems~\cite{okura2017embedding, liu2017related, wang2018billion, pal2020pinnersage, liu2022monolith} over multi-modal data.

In this work, we focus on the design and implementation of an embedded vector search engine for powering applications that run on users' personal devices. These personal devices are a rich source of information that very often cannot leave the device. Therefore, contextualized search and recommendation applications require vector search capability that can operate within the constraints of the device's environment (details in Section~\ref{sec:requirements}). Such use-cases can range from interactive sematic search over dynamic datasets to batch analytics workload involving structured attribute filters for constructing topically-related groups of assets. State-of-the-art vector data management systems~\cite{pan2024survey} generally targets scenarios where large servers with an abundance of memory are available; vector collections are static and inserts and deletes are not supported; and vector similarity search often happens in isolation of other search criteria such as filters over structured attributes. They lack the necessary optimizations to support scalable on-device vector search use-cases that need to operate in low-resource environments, while supporting updates and hybrid search queries that combine vector similarity search with structured attribute filters.

\begin{figure}
    \centering
    \includegraphics[width=1.0\linewidth]{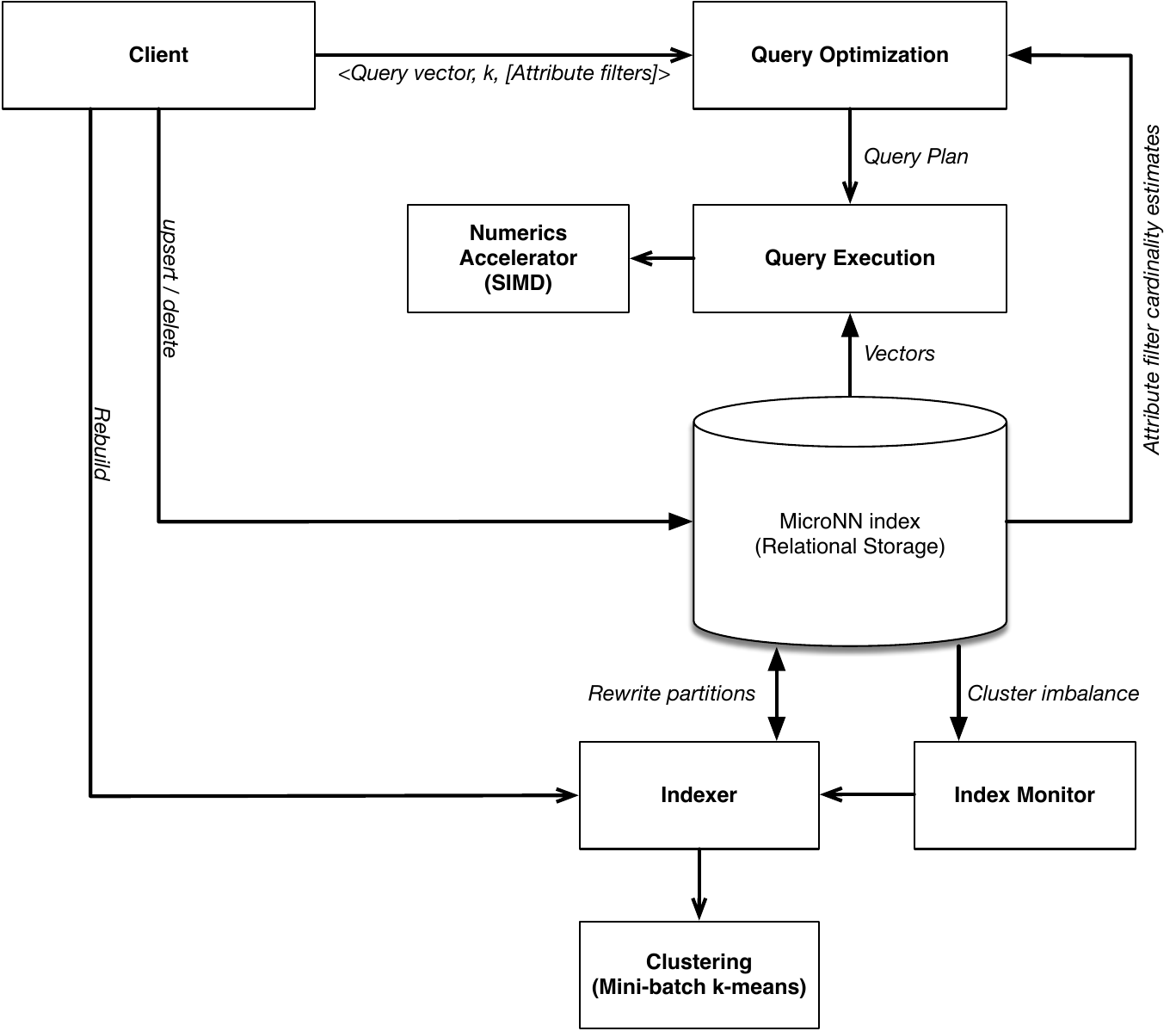}
    \caption{\micronn\ system architecture. Open arrows denote request flow, filled arrows denote read/write access.}
    \label{fig:micronn-arch}
\end{figure}

In this paper, we present \micronn, an embedded vector search engine that implements disk-resident vector indexing and vector search algorithms that are both memory efficient and performant (Figure~\ref{fig:micronn-arch}). To the best of our knowledge, \micronn is the first on-device vector search engine addressing the system and algorithmic challenges for supporting both exact K-nearest neighbour (KNN) and approximate K-nearest neighbour (ANN) search with structured attribute filters, and streaming inserts and deletes with ACID semantics for resource constrained environments. In particular, we make the following contributions:

\textbf{Disk-resident Vector Index and ANN Search:}
\micronn\ leverages an inverted-file-based (IVF) index that enables high-recall ANN search over large vector collections with interactive latency. It uses relational storage to ensure strong durability and consistency of the underlying data. The index is constructed using a memory efficient variation of the $k$-means clustering algorithm~\cite{sculley2010web} with balanced partition assignment~\cite{liu2018fast} to create clusters from the data vectors. At query time, the $n$ nearest clusters (based on distance from the query vector to the cluster centroids) are scanned to approximate the K nearest neighbours. This data partition pruning approach is parameterized to allow clients to trade-off latency for recall. One of our key contributions is the efficient movement of index partitions between disk and memory to use both resources effectively for striking a balance between memory usage and ANN search latency. In addition, we leverage SIMD accelerated floating point operations during query processing for performance, and implement efficient parallel heap structures for maintaining and merging result candidates during query processing. As a result of these optimizations, \micronn\ takes less than 7 ms for performing top-100 ANN search with 90\% recall on publicly available million-scale vector benchmark while using $\approx$10 MB of memory.

\textbf{Hybrid Search:}
\micronn\ supports structured attribute filters over user defined attributes that are combined with the vector similarity search by leveraging two query plans depending on attribute filter's selectivity. We have implemented a novel hybrid query optimizer that chooses between query plans for yielding the best ANN search latency while not sacrificing on search recall. 

\textbf{Batch Query Optimization:}
\micronn implements a batch query processing algorithm that incorporates multi-query optimization techniques in order to optimize I/O for query batches, giving significant gains in amortized latency per query as a function of batch size.

\textbf{Streaming Updates:}
\micronn\ supports streaming inserts (with upsert semantics) and deletes by leveraging a delta-store that is periodically incorporated into the main index. We employ an incremental IVF index update mechanism to reduce disk I/O when incorporating updates from the delta-store.

%% file: sections/motivation.tex
\section{On-device Vector Similarity Search}\label{sec:motivation}

\subsection{Use-cases and requirements}\label{sec:requirements}
Personal devices contain a wide variety of private information that can be used for contextualized search and recommendation. Very often user data needs to stay within user devices, necessitating the use of on-device indexing and search architectures for these search and recommendation applications. Furthermore, even for non-private data, on-device indexing can improve search latency and enable offline experiences even when network connectivity is poor.
As mentioned earlier, nearest neighbour search over dense vector embeddings is a crucial component of on-device search and recommendation systems, however, implementing efficient vector indexing and search on-device poses a number of unique challenges:
\begin{enumerate}
    \item \textbf{Resource-constrained environment:} User devices have varying capabilities; consequently, the deployed indexing and query processing algorithms must be capable of providing sufficient performance on environments with a wide range of memory and processing power. 
    \item \textbf{Multi-tenancy:} Hardware resources are shared across multiple applications, leading to strict memory constraints. Additionally, the index cannot be buffered in memory unless it is serving an active use-case, requiring indexing and search system to be disk-resident.
    \item \textbf{The need for IO efficiency:} Disk I/O is a crucial factor in both the performance of disk-based algorithms and the lifespan of flash-based storage devices, as high volumes of writes over time contribute to disk wear.
\end{enumerate}

In addition to the environment constraints above, on-device use-cases for vector similarity search can have a wide range of workload characteristics. 
Consider the following two personalized experiences built on vector similarity search:

\begin{example}[Interactive Semantic Search:]\label{ex:semantic}
Consider a scenario in which a large number of images on a mobile device have vector embeddings computed, and a user would like to search these images. First, retaining all of these embeddings in-memory is infeasible due to the burden this would put on the shared memory of the device. Second, as new images are added and others get deleted, the index needs to be maintained in real-time so that new images appear in searches and deleted ones do not. Furthermore, background processing on the image collection (\eg sync'ing inserts and deletes from the user's other devices) may produce concurrent reads and writes on the index. Third, the searches may be combined with structured attribute filters, such as date ranges or location. Finally, the search must have low-enough latency to be used in a real-time search experience, while having high enough recall to produce quality results.
\end{example}

\begin{example}[Visual Analytics]\label{ex:vu}
Analyzing images plays an important role in understanding the relationship between sets of assets in order to improve search or find related assets for constructing collections. A key part of visual analytics is in finding related items to a target asset. Such a workload would process many target assets in large batches to construct topically-related groups of assets. These searches may also involve structured attribute constraints, such as filters on timestamps or media types.
\end{example}

As exemplified above, supporting such on-device personalized experiences imposes the following requirements on the vector indexing and search algorithms:
\begin{enumerate}
    \item \textbf{Updatability:} The continuously changing nature of personal data on user devices requires the vector index to support dynamic updates. 
    \item \textbf{Consistency}:  Interactive on-device experiences that use embedding vectors for contextual search should reflect the current state of the system. Each reader should see a consistent state of the index at all times, including reading concurrently with writes and index maintenance operations.
    \item \textbf{Attribute predicates:}  Many applications of nearest neighbour search are done within the context of some other search criteria. As described above, on-device experiences often have additional metadata associated with each data item, and the similarity search needs to be done alongside other attribute constraints.
    \item \textbf{Diverse workloads:} In addition to providing interactive latencies, some workloads require high throughput when executing a large number of queries. The solution should support efficient execution of large query batches.
\end{enumerate}

\subsection{Limitations of Existing Methods}
\label{sec:limitations}

Despite the growing interest in systems and algorithms for vector similarity search (see \cite{pan2024survey} for a recent survey of vector database management systems), existing approaches face fundamental limitations that prevent their practical on-device deployment. We categorize these limitations into the following four key aspects and analyze how current systems are affected by them.

\subsubsection{Designed for abundant memory}

The varying capabilities and multi-tenant nature of user devices require that memory usage of any application must be constrained, and an index cannot be buffered in memory unless it is actively serving a use-case.
However, designed for cloud-based deployments for web-scale datasets in mind, the majority of existing approaches rely heavily on the availability of abundant memory and target main memory indices. 
For instance, most LSH-based methods require substantial memory due to their inherent need for high redundancy \cite{sundaram_streaming_2013, zheng2020pm}. Similarly, commonly used vector similarity search libraries such as FAISS\cite{johnson2019billion} and HNSWlib\cite{noauthor_hnswlib} are main memory only implementations. 
Based on the assumption that all vector can be stored in memory, these approaches focus on minimizing the number of distance comparisons. Disk residency also requires optimizing the number of disk accesses.
DiskANN~\cite{jayaram2019diskann} is one exception that provides SSD-based vector indices. However, it still requires a compressed version of all vectors to be buffered in main memory which is not suitable for our scenarios.

\subsubsection{Lack of updatability \& consistency guarantees}

Current systems predominantly optimize for read-only workloads, and the continuously changing nature of personal data on user devices poses challenges to existing vector similarity search approaches. 
Tree-based methods~\cite{annoy, kd-tree} suffer from quality degradation due to imbalances as existing vectors are deleted and new vectors are inserted. Similarly, graph-based methods such as HNSWlib~\cite{noauthor_hnswlib} and NSG~\cite{fu12fast} are negatively affected due to perturbations in the graph structure, and they incur high computational costs due to random access required by graph modifications. Partition-based techniques (such as FAISS-IVF\cite{johnson2019billion}) are amenable to efficient updates. However, as vectors are added, deleted, or updated, the pre-computed centroids no longer reflect the state of vectors in the clusters. As re-clustering the data on every insert is infeasible due to the high latency of the clustering algorithms and the amount of I/O it generates, most existing systems rely on periodically rebuilding the entire index. Although it can be acceptable for applications with static or slowly changing datasets, it can result in stale search results as the updates between rebuilds are not reflected in the index.

FreshDiskANN\cite{singh2021freshdiskann} and SPFresh\cite{xu_spfresh_2023}, notable exceptions that support real-time updates, are able to handle concurrent reads and writes without significant degradation in index quality. However, their memory-resident index structures makes them impractical for our on-device scenarios where the index size is significantly larger than the available memory.

\begin{table*}[t]
	\caption{An overview of existing approaches for vector indexing and search, and characteristics based on the requirements from Section \ref{sec:requirements} }\label{tab:existing}
    \small
    \begin{tabular}{| c| c| c| c| c| c| c|}
		\hline
        Type & Name & Constrained memory & Updatability & Consistency & Hybrid queries & Batch queries \\ 
		\hline
		\hline
        \multirow{3}{*}{LSH}
        & PLSH~\cite{sundaram_streaming_2013} & $\times$ & $\checkmark $ & $\checkmark$ & $\times$ & $\times$ \\
        & PM-LSH~\cite{zheng2020pm} & $\times$ & $\checkmark $ & $\checkmark $ & $\times$ & $\times$ \\
        & HD-Index~\cite{10.14778/3204028.3204034} & $\checkmark $ & $\checkmark $ & $\checkmark $ & $\times$ & $\times$ \\
        \hline
        \multirow{2}{*}{Tree} 
        & kd-tree~\cite{kd-tree} & $\times$ & $\checkmark $ & $\checkmark $ & $\times$ & $\times$ \\
        & Annoy~\cite{annoy} & $\checkmark $ & $\checkmark $ & $\checkmark $ & $\times$ & $\times$ \\
        \hline
        \multirow{3}{*}{Graph} 
        & HNSWlib~\cite{noauthor_hnswlib} & $\times$ & $\times$ & NA & $\times$ & $\times$ \\
        & DiskANN~\cite{jayaram2019diskann, singh2021freshdiskann} & $\times$ & $\checkmark$ & $\times$ & $\checkmark$ & $\times$ \\
        & ACORN~\cite{patel2024acorn} & $\times$ & $\times$ & NA & $\checkmark$ & $\times$ \\
        \hline
        \multirow{4}{*}{Partitioned} 
        & FAISS-IVF~\cite{johnson2019billion} & $\times$ & $\times$ & NA &  $\checkmark$ & $\checkmark$ \\
        & Milvus~\cite{wang2021milvus} & $\times$ & $\checkmark $ &  $\checkmark $ &  $\checkmark $ & $\times$ \\
        & SPANN~\cite{chen_spann_nodate} & $\checkmark$ & $\times$ & NA & $\times$ & $\times$ \\
        & SP-Fresh~\cite{xu_spfresh_2023} & $\checkmark$ & $\checkmark$ & $\checkmark$ & $\times$ & $\times$ \\
        & \textbf{MicroNN} & $\checkmark $ & $\checkmark $ & $\checkmark $ & $\checkmark $ & $\checkmark $ \\
        \hline
  	\end{tabular}
\end{table*}

\subsubsection{Batch query optimizations}

In addition to providing individual search latencies for interactive applications, on-device vector similarity search algorithms should support efficient execution of large query batches, such as performing concurrent KNN or ANN search for multiple vectors, or computing distances against all vectors in the index. 
Such batch settings provide opportunities for multi-query optimization. However, to the best of our knowledge, most existing systems aim to reduce the latency of individual online queries and do not optimize for throughput. 
In particular, tree- and graph-based indices are inherently challenging for multi-query optimization due to graph traversal-based search algorithms. Hence, graph-based indices such as HNSWlib~\cite{noauthor_hnswlib} and DiskANN~\cite{jayaram2019diskann} default to processing each query in a batch individually without providing multi-query optimizations.
In contrast, partitioned data layout in IVF-based indices present opportunities for computation sharing and parallelization. HQI~\cite{mohoney_high-throughput_2023} first introduced multi-query optimizations for vector similarity search. However, it does not support updates as it is designed for static, memory-resident workloads.

\subsubsection{Limited support for hybrid queries}
Although the primary focus of most vector data management system is to optimize vector similarity search, many industrial applications of nearest neighbour search are done within the context of some other search criteria.

As an example, consider the scenario in Example \ref{ex:semantic}, and assume that the user has 100,000 photos in their image collection. If the user lives in New York, but once took a trip to Seattle and took 15 photos there, then a mere 0.015\% of the photos will satisfy the  predicate \textit{location = ``Seattle''}. This is called a ``high selectivity predicate''. If, on the other hand, the user lives in Seattle and 95\% of their photos match the predicate \textit{location = ``Seattle''}, then this is called a ``low selectivity predicate''. The ideal query plans for high and low selectivity predicates are very different, and running the wrong query plan can result in very long search latency or very low recall (often with empty result sets). In addition, most real-world applications exhibit queries with complex attribute constraints; yet, the support for attribute constraints in existing systems~\cite{gollapudi2023filtered, wang2021milvus} are limited to numerical comparisons or exact text matches.

\subsection{Capabilities of Existing Approaches}
Table~\ref{tab:existing} summarizes the capabilities of existing approaches compared to \micronn. 
The ``Constrained memory'' column indicates whether the approach can run with less memory than the size of the index (or conversely, if it requires the index to be fully loaded into memory for query processing.) ``Updatability'' indicates whether or not the approach allows updates without fully rebuilding the index. 
``Consistency'' indicates whether or not the approach guarantees consistent snapshots under concurrent reads and writes (\ie transaction isolation.) The ``Hybrid queries'' column shows if the approach supports nearest neighbour queries with structured attribute constraints. Finally, the ``Batch queries'' column indicates whether or not the approach supports a batch query interface.

While some approaches can run in constrained memory environments, they do not support updates, hybrid queries, and batch queries. Other approaches support hybrid queries and batch query interfaces, but rely on the index being fully loaded into memory. All of these capabilities are required in order to support the wide range of workloads described in Section~\ref{sec:requirements}.

%% file: sections/overview.tex
\section{\micronn}\label{sec:overview}

We introduce \micronn (Micro Nearest Neighbours), an ANN search system purpose-built for on-device vector similarity search. \micronn indexing and search algorithms are \textit{disk-resident}, and efficiently and seamlessly utilize SSD storage to satisfy strict memory constraints, making it suitable for on-device deployments. In addition, \micronn's indexing algorithms supports \textit{real-time updates}, and introduces optimizations for \textit{batch query processing} and \textit{hybrid-query processing}. \micronn\ is implemented as a software library that can be linked by any application to create their own local vector index. 

Figure \ref{fig:micronn-arch} shows the architecture of \micronn.  
We adopt a relational storage architecture and leverage a SQLite relational database for efficient storage of vectors and their associated metadata. 
For indexing, we implement an Inverted File (IVF) index~\cite{jegou_product_2011}, where the vector space is partitioned into clusters using a variation of k-means algorithm optimized for low-resource environments~\cite{sculley2010web}. The resulting clusters constitute the partitions of the index. By storing vectors in a clustered manner on disk, \micronn aims to improve locality and I/O efficiency when accessing a partition. Additionally, \micronn supports real-time updates using a delta-store, enabling concurrent reads and writes. The use of a relational data store also enables index rebuilds to be performed concurrently with transactionally consistent reads. An index monitor is used to track index quality upon updates, and triggers re-indexing when necessary.

\micronn also implements a simple query optimizer to find an efficient execution strategy for hybrid queries which combine ANN search with structured attribute filters.
For these \emph{hybrid queries}, the query optimizer relies on selectivity estimations for choosing between pre- and post-filtering query execution plans.

For batch workloads, we implement a multi-query optimization technique to amortize partition scan costs and reduce I/O. 

In the following section, we describe the components of \micronn in detail and highlight the contributions that enable \micronn to address the challenges and requirements of on-device deployments described in Section~\ref{sec:motivation}.

%% file: sections/indexing.tex
\subsection{Vector Indexing}\label{sec:indexing}

Inverted File (IVF) is a commonly used index structure in vector databases, where a vector quantization algorithm (\eg k-means) is used to partition a set of vectors. Such clustering results in similar vectors being assigned to the same cluster and these clusters are used as the data partitions of the index~\cite{jegou_product_2011}.

Despite widespread adoption of IVF indexes in vector databases,  \cite{wei2020analyticdb, ferrari_revisiting_2018, jégou2011searching, guo_accelerating_2020, chen_spann_nodate, babenko_inverted_2015}, out-of-the-box IVF implementations are not suitable for on-device deployments due to two main issues. 
First, the most of the quantization algorithms used for clustering (the most typical one being k-means) require the entire vector set to be buffered in memory, which is not feasible for on-device workloads where memory is constrained.
Second, the clustering algorithm may produce clusters of various sizes. As shown in \cite{mohoney2024incremental}, partition imbalance is an indicator of query performance for partitioned indexes, \textit{i.e.}, uneven size distribution of partitions adversely affects the query processing performance.

To address the aforementioned issues, we implement a variation of the k-means algorithm optimized for resource-constrained environments. In brief, \micronn's indexing algorithm is based on mini-batch k-means~\cite{sculley2010web} for reducing memory footprint of indexing, and it implements flexible balance constraints to reduce variation in partition sizes \cite{liu2018fast}.

\begin{algorithm} 
\caption{\micronn Clustering Algorithm} \label{alg:indexing}
\KwIn{Input vector set $\mathbf{X}$, target cluster size $t$, mini-batch size $s$, number of iterations $n$}
\KwOut{Set of centroids $\mathbf{C}$, cluster assignment $\mathbf{P}: \mathbf{X} \rightarrow \mathbf{C}$}
\SetKwBlock{Begin}{function}{end function}
    $k \leftarrow |\mathbf{X}| \big/ t$ \tcp{\# of clusters}
    $C \leftarrow \{c_0, c_1, \cdots c_{k-1}\}$ \tcp{Initialize each cluster centroid c with a random $\mathbf{x}$ $\in$ $\mathbf{X}$}
    $\mathbf{v} \leftarrow \emptyset$ \tcp{Store cluster sizes}
    $\mathbf{d} \leftarrow \emptyset$ \tcp{temp cluster assignments}
    \For{$i = 1$ \textnormal{to} $n$} {
        $M \leftarrow s$ examples picked randomly from $\mathbf{X}$
        
        \For{$\mathbf{x} \in \mathbf{M}$} {
            $d[\mathbf{x}] \leftarrow NEAREST(\mathbf{C}, \mathbf{v}, \mathbf{d}, \mathbf{x})$ \tcp{find the nearest centroid for $\mathbf{x}$ within balance constraints}
        }
        \For{$\mathbf{x} \in \mathbf{M}$} {
            $c \leftarrow d[\mathbf{x}]$ \tcp{Retrieve cached center for $\mathbf{x}$}
            $v[c] \leftarrow \mathbf{v}[c] + 1$ \tcp{Update per-center counts}
            $\eta \leftarrow \frac{1}{\mathbf{v}[c]}$ \tcp{Compute per-center learning rate}
            $c \leftarrow (1 - \eta)\mathbf{c} + \eta \mathbf{x}$ \tcp{Update centroid position}
        }
    }
    $P \leftarrow 0$ \tcp{Initialize partition assignments}
    \For{$\mathbf{x} \in \mathbf{X}$} {
        $\mathbf{P}[\mathbf{x}] \leftarrow g(\mathbf{C}, \mathbf{x})$ \tcp{Assign nearest center to $\mathbf{x}$}
    }
    \Return $\mathbf{C}$ (centroids) \&
    $\mathbf{P}$ (partitions)
\end{algorithm}

Algorithm \ref{alg:indexing} outlines the clustering used during indexing in \micronn. First, the value of $k$, (the number of clusters) is determined based on the size of the vector dataset and the target cluster size. This can be tuned per use case, by default we target 100 vectors per cluster.
During each iteration of  Algorithm \ref{alg:indexing}, cluster centroids are computed over a batch of size $\mathbf{s}$ that is uniformly randomly sampled from $\mathbf{X}$ \cite{sculley2010web}. By performing cluster assignments in small batches, \micronn eliminates the need to buffer all vectors in memory.
In order to ensure balanced partitioning, the function $NEAREST$, which finds the nearest centroid for a given vector $X$,  uses a penalty term for large clusters. As such, the resulting index has more balanced clusters, and vectors are spread out among nearby clusters instead of creating a few ``mega'' clusters.

During clustering, Algorithm \ref{alg:indexing} computes the similarity between vectors and the cluster centroids, which is a CPU intensive operation for high-dimensional vectors. We optimize vector similarity computations during index construction by representing a batch of vectors as a matrix and use an hardware accelerated linear algebra library that utilizes SIMD operations.

After Algorithm \ref{alg:indexing} produces the final clustering, cluster centroids are persisted and the vectors' partition assignments are updated in the underlying database.

\subsection{Physical Storage}\label{sec:physical-storage}
\begin{figure*}
\includegraphics[width=1.0\textwidth]{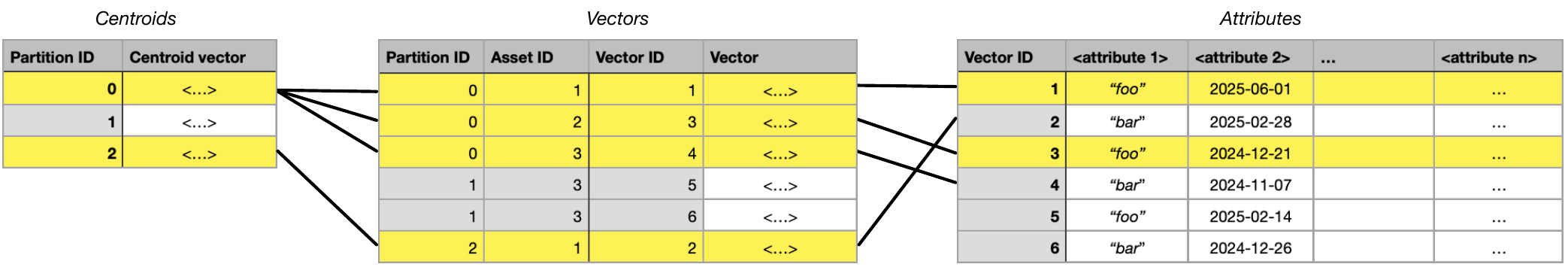}
\caption{Relational schema and example scan result. Centroid vectors are scanned to find the $n$-nearest centroids to the query vector (partition 0 and 2). The corresponding vectors with partition 0 or 2 are scanned from the \textit{Vectors} table, which is clustered on \textit{Partition ID} for data locality. The \textit{Vectors} table is joined with the \textit{Attributes} table, which has a filter constraint on the user defined indexed attribute \textit{attribute 1} = \textit{``foo''}. Vectors $1$ and $3$ are returned as the nearest neighbours that also satisfy the attribute constraint. }
\label{fig:example-schema}
\end{figure*}

\micronn is designed as \textit{middleware} for on-device vector indexing and search, and uses a relational database for storage. We opt to use SQLite as it is a self-contained, reliable, small relational database engine with strong durability and isolation guarantees. Using an existing relational database engine for physical storage has a number of benefits:
\begin{itemize}
    \item \textbf{concurrency}: building on SQLite's concurrency control, \micronn allows concurrent clients: a single writer (performing upserts, deletes, and index rebuilds) and multiple readers across threads and processes;

    \item \textbf{performance}: SQLite provides scan throughput on-par with native file access; 

    \item \textbf{clustering support}: the IVF index layout can easily be reflected by using a clustered primary index;

    \item \textbf{maturity}: \micronn inherits proven durability, isolation, and recoverability of its data artifacts without needing to re-invent a transactionally consistent storage layer to underpin the vector index.
\end{itemize}

\micronn directly manages the underlying SQLite database in order to control how the vectors and their metadata are stored on disk. 
In particular, \micronn stores the vectors as blobs in a relational table, with the partition ID (given by the IVF clustering), asset ID (given by the client to denote which asset produced this vector), and vector ID (generated internally) as the primary key. A clustered index ensures that the rows of the vector table are clustered on disk, giving data locality to vectors in the same partition. After a (re)clustering operation is performed by the index construction process, the partition IDs in the vector table are updated. 

Centroids are stored in a separate table. This table is significantly smaller than the vector table and can be scanned to find the nearest centroids to the query vector. To scale to even larger collections, the centroid table itself could also be indexed. We have found this to be unnecessary for the workloads \micronn currently supports.

The use-case specific attributes are stored in a separate attribute table. Each vector can have its own attribute values, and nearest neighbour queries can include relational constraints over these attributes (see Section~\ref{sec:search}).

Figure~\ref{fig:example-schema} shows the relational tables for an example \micronn database. The IVF clustering performed by Algorithm \ref{alg:indexing} (Section \ref{sec:indexing}) in this example produces three clusters. The centroid vectors for these clusters, along with an auto-generated partition ID are stored in the \textit{Centroids} table. Each vector stored in the \textit{Vectors} table uses the partition ID as the clustering key, providing data locality on disk for the vectors that belong to the same partition.

%% file: sections/query.tex
\subsection{Vector Similarity Search}\label{sec:search}

\micronn supports both exact K-nearest neighbour (KNN) and approximate K-nearest neighbour (ANN) search. The former is a trivial but resource intensive operation as it requires exhaustive scan of the entire Vectors table. ANN search, on the other hand, enables \micronn to control the trade-off between query latency and recall (the percentage of vectors in the approximate top-K present in the exact top-K vectors). Here, we describe the search algorithm adopted in \micronn, followed by the query optimizations \micronn implements. 

\micronn adapts the traditional IVF search algorithm to include delta-store scans and utilizes a number of engineering optimizations to ensure run-time performance. The ANN search algorithm is parameterized by a query vector $q$, query limit $k$, and a number of partitions to scan $n$. As shown in Algorithm \ref{alg:ann-search}, vector search operation is optimized through a two-level approach instead of comparing the query vector against every vector in the database. First, the list of cluster centroids are scanned and and their distance to the query vector is calculated.
Then, Algorithm \ref{alg:ann-search} selects the $n$ nearest partitions based on the distance between the query vector and centroid. The similarity between the query vector and all data vectors in this partition are computed while scanning each partition, and the top $k$ nearest vectors are maintained in a heap. The delta partition is always included in addition to the $n$ selected partitions. This ensures that any newly inserted vectors are also considered in the result.
This significantly reduces the number of distance computations compared to exhaustive search, while still maintaining high accuracy by focusing on the most relevant partitions. Finally, the algorithm returns the $k$ most similar vectors sorted by their similarity scores.
\todo{including vector batching in algorithm}
\begin{algorithm} 
\caption{\micronn ANN Search Algorithm} \label{alg:ann-search}
\KwIn{Query vector $q$, limit $K$, \# of partitions to scan $n$}
\KwOut{$K$ nearest vectors and their distances to $q$} 
\SetKwBlock{Begin}{function}{end function}
    $\mathbf{P} \leftarrow centroids$ \tcp{cluster centroids form the index}
    $\{\mathbf{R}\} \leftarrow \emptyset $ \tcp{initialize a set of max heaps, one for each worker thread} 

    $\mathbf{C} \leftarrow$ FindNearestCentroids($\mathbf{P}$, $q$, $n$) $\cup$ deltaPartition\\
\tcp{Parallel iteration over $n+1$ partitions}
    \For{$\{c_i\} \in \mathbf{C} $}{ 
     \tcp{scan the clusters}
        \For{$v_j \in \mathbf{Scan}(c_i)$ }{
            $d_i$ = ComputeDistance($v_j$, $q$) \\
            \tcp{maintain top-$K$}
            \If{ $\mathbf{R}_i$.{size()} < $K$}{
                $\mathbf{R}_i$.Enqueue($v_j$, $d_i$)
            }
            \If{$d_i$ < $\mathbf{R}_i$.Peek().distance}{ 
                $\mathbf{R}_i$.replaceMax($v_j$, $d_i$)
            }
        }
    }

    \Return Sort($\cup_i$ $\mathbf{R}_i.vectors$, $\mathbf{R}_i.distances$)
\end{algorithm}

A higher value of $n$ provides higher recall as more data is scanned, but with longer latency. An optimal value for $n$ depends on the workload, data distribution, and database size.

\begin{figure}
    \centering
    \includegraphics[width=0.925\linewidth]{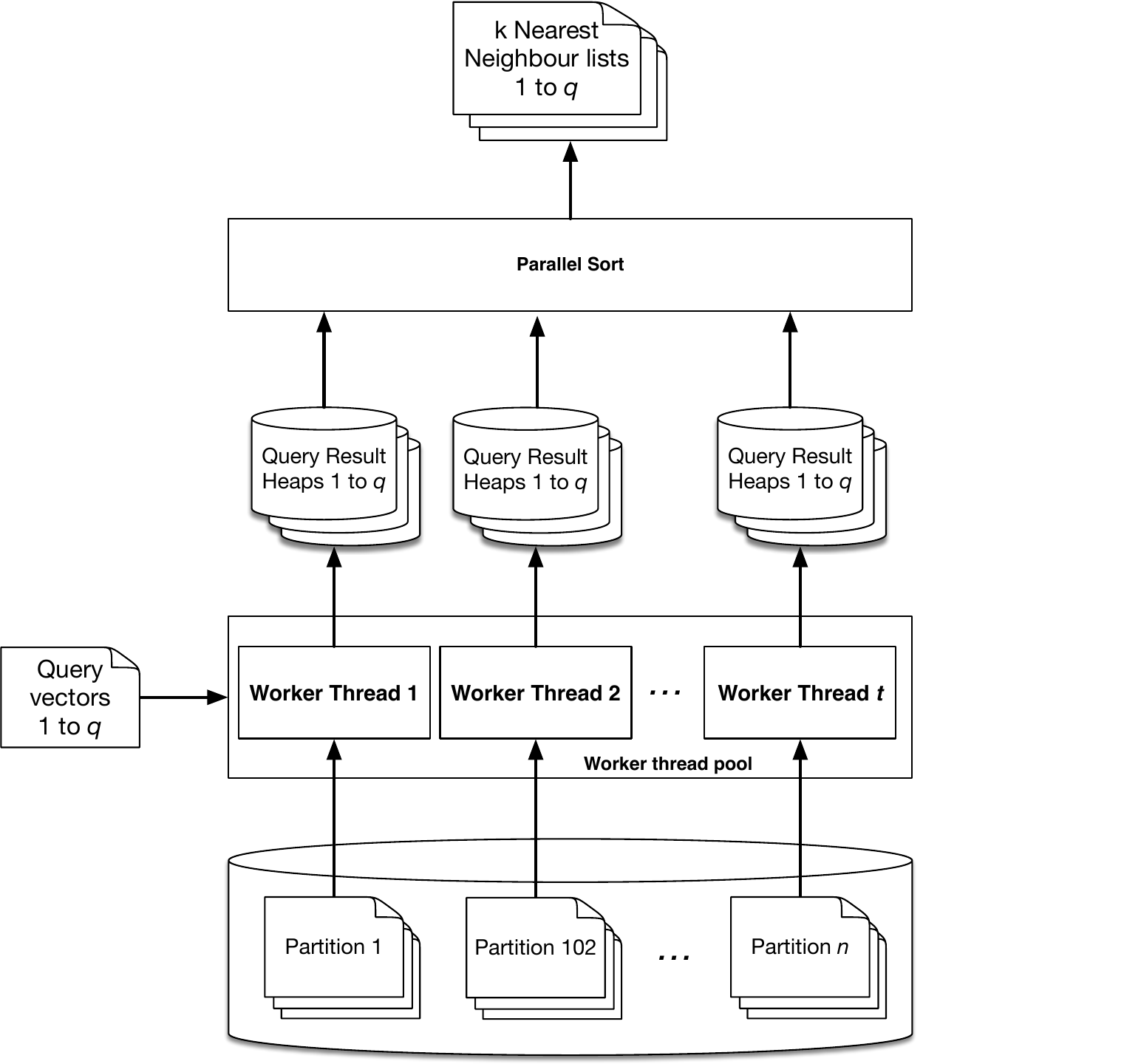}
    \caption{Query processing pipeline}
    \label{fig:query-pipeline}
\end{figure}

To accelerate query processing, \micronn also includes a number of engineering optimizations. First, data partitions are scanned in parallel to ensure we fully utilize the available disk bandwidth. The distance calculations are assigned to a number of threads in order to leverage mutli-core CPUs. Each thread maintains its own heap of its current top-$k$ vectors, and an efficient parallel heap merge is performed once all threads finish processing their partitions. Finally, distance computations are done over batches of vectors. Each vector in a  batch is inserted into a matrix where SIMD operations can be leveraged to parallelize the query to vector distance calculations. By storing the vector blobs in the database using the format expected by the matrix multiplication library, we eliminate expensive data marshalling operations and minimize the number of copy operations performed on the vectors. Figure~\ref{fig:query-pipeline} illustrates the query processing pipeline, including concurrent processing for $q$ query batches which we describe in Section~\ref{sec:batch}

\subsection{Batch Query Processing}\label{sec:batch}

In addition to interactive ANN search, on-device analytics workloads such as image analyses described in Example~\ref{ex:vu} require high throughput processing of a batch of queries.
A na\"\i ve approach to processing a batch of queries is to simply dispatch the queries concurrently, but process them independently.
Although functionally correct, such an approach results in unnecessary computation as multiple queries might re-scan the same partition.
Relational databases have long implemented approaches to multi-query optimization (MQO). Multi-query optimization reuses data artifacts required by concurrently running queries (base table scans, index scans, or even intermediate results).
In our prior work we demonstrated that MQO can significantly improve batch ANN query processing the throughput over partitioned indexes such as IVF~\cite{mohoney_high-throughput_2023}.

Inspired by HQI~\cite{mohoney_high-throughput_2023}, \micronn implements a variation of MQO for processing a batch of queries. 
Given a batch of queries, \micronn first identifies the set of clusters that each query needs to access, and groups queries per partition.
 Then, instead of scanning a partition multiple times for each query, distances between queries and the vectors in the partition is calculated via a single matrix multiplication.
This amortizes the cost of scanning a partition over a batch queries; greatly reduces I/O, and yields significant improvements in amortized aggregate throughput.
Based on our empirical evaluation on an internal analytics workload for a recommendation use-case, individual query latency is cut down by more than 30\% when queries are executed in batch of 512 queries.

\subsection{Hybrid Query Support}
\micronn supports processing of \textit{hybrid queries} \cite{mohoney_high-throughput_2023} by allowing user-defined attributes to be stored along side the vector data and attribute constraints in the form of relational predicates to be applied. 
For example, find the 10 most similar assets to the embedding of the phrase \textit{''black cat playing with yarn''} with the constraint  \textit{location = ``Seattle''}. In this example, the user of \micronn would have defined location as a filterable attribute and \micronn supports standard relational operators over the defined attributes (>, <, =, !=). Client defined attributes are indexed using sqlite’s b-tree implementation making search over attributes efficient. In addition, \micronn allows a full-text index (FTS) to be created over filterable attributes. Clients can combine nearest neighbour search with text search using SQLite’s FTS5 search syntax. 

\micronn implements two different algorithms for nearest neighbour search combined with attribute filtering. 

\textit{Post-filtering} first processes the nearest neighbour search computing up to k results. The results are then joined against the Attributes table with the constraints applied to filter the result list. This approach is efficient (a small operation after the  ANN search), but may affect recall as the result list is filtered. An important optimization here is that when the IVF partitions with vector blobs are retrieved from the database, we apply the join and filter over the Attributes table. Vectors in the requested partitions that don't satisfy the predicate filter are therefore filtered before being considered in the top-K computation.

\textit{Pre-filtering} evaluates the predicate filter before conducting the vector similarity search. From the Attributes table, we evaluate the attribute filter and produce a set of matching asset ids. For every asset id that satisfies the attribute filter, the algorithm fetches the vectors from the Vector table, computes similarity against the query vector, and maintains the top-K results in a heap. Pre-filtering scans all vectors that satisfy the attribute filter.  This is a brute force nearest neighbour search over the subset of the data qualified by the filter which guarantees 100\% recall. However, the latency of the pre-filtering depends on how many results satisfy the filter, also known as \emph{predicate selectivity}.

\subsubsection{A Query Optimizer for Hybrid Queries} \label{sec:queryoptimizer}

Given these two query plans, we can reason about choosing one or the other by computing estimates of selectivity \cite{selinger1979access}. Selectivity Factor $F$ is the ratio of rows which are qualified by a given predicate to the total size of the relation $R$:

\begin{equation}
F = \frac{|\sigma_{predicate}(R)|}{|R|}
\end{equation}

We say a predicate is \emph{highly selective} when it has a low selectivity factor (\ie it qualifies very few rows). A predicate has \emph{low selectivity} when it has a high selectivity factor (\ie it qualifies many rows).

If a predicate has a low selectivity factor then pre-filtering is the optimal query plan. It has 100\% recall, and since very few results are returned from the filter, the brute force nearest neighbour search is efficient. Revisiting our \emph{``black cat playing with yarn''} in Seattle query example, if someone took a small number of photos when visiting Seattle and uses that location as a filter during search, then that predicate will be highly selective. It will qualify only a small fraction of the photos out of the entire photo library. In this scenario, pre-filtering is very efficient and would yield 100\% recall.

If a predicate has a high selectivity factor then post-filtering is the optimal query plan. Pre-filtering would be inefficient since it will perform brute force nearest neighbour search over the qualified results, and high selectivity factor means a large result set. Post filtering on the other hand, will use ANN search then discard any results which do not satisfy the predicate. The number of results discarded is proportional to the selectivity of the predicate. Continuing the example from above, when the attribute filter \textit{location = ``Seattle''} is applied over a photos collection where the user lives in Seattle and 95\% of the photos qualify for the predicate, then the predicate has very low selectivity as it will qualify most of the photos.

\micronn implements a simple query optimizer based on predicate selectivity estimates. Different from a traditional relational optimizer, the choice of query plan for vector search affects recall as well as latency. 
We can view the IVF search as a predicate that filters the partition id column, and estimate its selectivity based on the number of partitions we will scan and each partition's target size. Given the number of IVF partitions $n$ and a target partition size $p$, an estimate of the selectivity factor of our IVF predicate filter $\hat{F}_\text{IVF}$ is:

\begin{equation}
\hat{F}_\text{IVF} = \frac{n \cdot p}{|R|} \label{eq:scan_estimate}
\end{equation}

If we estimate that the selectivity of the actual attribute filter is below $\hat{F}_{IVF}$, it's better to apply pre-filtering, as the attribute filter itself narrows our search space more than the IVF index. If our estimated selectivity is above $\hat{F}_\text{IVF}$, the IVF index constrains the search space more the attribute filter, and post-filtering is the better strategy. This optimization strategy ensures the latency of the query will match either pre-filtering or post-filtering. It is always possible for a client to dynamically change the number of probes to increase recall at the expense of latency.

To estimate selectivities, we estimate the cardinality of applying each predicate of the attribute filters independently. For simplicity, we assume independence of the predicates and take the minimum over conjunctions and a sum over disjunctions to estimate the total selectivity of all attribute filters together. 
We denote the final cardinality derived from this procedure as $|\hat{\sigma}_{filters}(R)|$. Our selectivity factor estimate $\hat{F}_{filters}$ for all the attribute filters is then:

\begin{equation}
\hat{F}_{filters} = \frac{\min(|\hat{\sigma}_{filters}(R)|, |R|)}{|R|}
\end{equation}

The query optimizer then selects the pre-filtering query plan if $\hat{F}_{filters} < \hat{F}_\text{IVF}$, otherwise it selects post-filtering. We recognize that the wealth of research in query optimization and selectivity estimates could be leveraged to further improve the optimizer, and leave exploration of more sophisticated optimization techniques for future work.

\subsection{Updates}\label{sec:updates}
\micronn supports inserts (with “upsert” semantics in case the asset ID already exists in the database) and deletes. \micronn supports multiple readers scanning the IVF index in a snapshot isolated way. The writes, including adding, updating, and deleting assets, as well as index rebuilds are fully serialized. \micronn configures the underlying SQLite database to operate in write-ahead logging (WAL) mode for enabling ACID properties and protecting the stored vectors from corruption.

Inserts are processed by inserting into a “delta-store”. The delta-store is a set of recent vectors that have not yet been assigned a partition. Newly inserted vectors are staged in the delta-store until the IVF index is rebuilt. The delta-store is fully scanned on every query. This means that query latency can grow if the delta-store grows too large, therefore, requires periodic index rebuilds to flush delta-store content into the IVF index. 

While delta-store is logically different from the IVF index, in our implementation of \micronn, the delta-store is physically co-located with the IVF index. The delta-store is represented by assigning a reserved partition identifier, in this way vectors assigned to the delta-store can be represented in the same way as vectors assigned to the IVF index partitions. Adopting the same physical layout also brings in the advantage of ensuring data locality of the vectors assigned to the delta-store. Therefore, during nearest neighbour search, the delta-store is simply an additional partition.

A key challenge in incorporating the changes in delta-store into the IVF index is to do that in a way that does not block interactive applications, and also ensure unnecessary I/O is avoided to protect solid-state storage devices commonly found in user devices. In a separate work, we dive deep into the design space of what metrics can be tracked for assessing the query recall and latency characteristics of an IVF index, and what actions can be taken to mitigate degradation of these metrics over time~\cite{mohoney2024incremental}. For implementation purposes, we use a simplified form of incremental index maintenance that flushes  vectors from the delta-store by assigning them to the IVF index partition with the closest centroid and updates the centroids to reflect the partition content~\cite{arandjelovic2013all}. This approach of vector assignment can lead to partition sizes growing, consequently, leading to increased query latency. We prevent unbounded growth of query latency by allowing clients to put a threshold on average partition size growth. When average partition size reaches its growth limit, \micronn\ will trigger a full index rebuild. 

%% file: sections/experiments.tex
\section{Empirical Evaluation}\label{sec:experiment}
We empirically evaluate \micronn using both publicly available and internal workloads and compare against a number of different approaches for ANN search. We first describe our experimental setup, we present end-to-end evaluation of \micronn followed by a number of micro benchmarks evaluating specific technical contributions. The highlights of the evaluation are:
\begin{itemize}
    \item \micronn can perform ANN search with latency on-par with a full in-memory indexing approach while using two orders of magnitude less memory.
    \item \micronn uses 4$\times$ - 60$\times$ less memory during index construction compared to constructing a k-means based IVF index while maintaining similar index quality.
    \item For queries with attribute predicates, \micronn's optimizer can find an efficient execution plan by estimating predicate cardinality using per-column histograms.
    \item \micronn exhibits sub-linear scaling \textit{w.r.t.} batch size. 
\end{itemize}
\subsection{Experiment Setup}

\subsubsection{Datasets}
We use a number of publicly available datasets with sizes varying from tens of thousands to millions of vectors of varying dimensionality. We also use an internal workload used for a search and recommendation use-case containing approximately 150k 512-dimensional training vectors and 1000 test vectors. Characteristics of the datasets used are described in Table~\ref{tab:dataset}. 

\begin{table}
    \small
    \centering
    \caption{Datasets used in the evaluation}
    \begin{tabular}{|c|c|c|c|c|}
        \hline
        Dataset & Dimension & Vectors & Queries & Metric \\
        \hline
        \hline
        MNIST & 784 & 60k & 10k & L2 \\
        \hline
        NYTimes & 256 & 290k & 10k & cosine \\
        \hline
        SIFT & 128 & 1M & 10k & L2 \\
        \hline
        GLOVE & 200 & 1.18M & 10k & L2 \\
        \hline
        GIST & 960 & 1M & 1k & L2 \\
        \hline
        DEEPImage & 96 & 10M & 10k & cosine \\
        \hline
        InternalA & 512 & 150k & 1k & cosine \\
        \hline
    \end{tabular}
    \label{tab:dataset}
\end{table}

\subsubsection{Execution Environment}
\micronn is designed for edge environments with limited compute and memory. We run the experiments on multiple edge devices representing commonly used personal and portable computing devices. These devices have one or more CPU cores and memory capacity varying from a few gigabytes to a few tens of gigabytes. In the following we use \textbf{Small} and \textbf{Large} to refer to device under test (DUT) with single digit and a few tens of Gigabyte of main memory, respectively. 

\subsubsection{Evaluation Metrics}
Our evaluation is primarily focused on measuring the following:
\begin{itemize}
    \item \textbf{Query processing latency:} the time needed to perform top-100 nearest neighbour search at 90\% recall for a dataset.
    \item \textbf{Index construction time:} the time needed to construct an IVF index from scratch.
    \item \textbf{Memory usage:} total memory usage during (i) query processing, and (ii) index construction.
\end{itemize}
Unless otherwise specified, we report query processing latency and memory usage for finding top-100 nearest neighbours at 90\% recall. 
In most experiments, we use an average cluster size of 100 when building the index for each dataset. We then identify $n$, the number of IVF index partitions to scan to reach a recall of 90\% or higher.  We report the average value of the metrics over the query set and use error bars to donate standard deviation where applicable.

\subsubsection{Compared Approaches} We compare our proposed approach against a main memory-only variation of \micronn in order to keep all implementation aspects fixed and evaluate the effect of a disk-resident index. We also evaluate \micronn against short- and long-lived application patterns to demonstrate the impact of cold start. Details of the compared approaches are as follows:
\begin{itemize}
    \item \textbf{InMemory:} A completely memory resident variation of the \micronn IVF index. This baseline gives a lower-bound on latency for our IVF implementation, while illustrating the memory requirements to achieve this latency.
    \item \textbf{\micronn-ColdStart:} \micronn where all cached disk pages are purged from memory before running the benchmarks, and measurement is taken only for a single query. This variation demonstrates the impact of cold database cache. We repeat the evaluation for 100 randomly sampled queries from benchmark datasets and report the mean of the metric values along with standard error. This scenario represents short-lived applications or an application's bootstrap scenario. 
    \item \textbf{\micronn-WarmCache}: \micronn where the database caches are pre-warmed by running batches of queries prior to taking measurement of the subsequent benchmark. This scenario represents the commonly found pattern in long-lived applications that persist database connections.
\end{itemize}

We evaluate all three scenarios for demonstrating the end-to-end performance. 
The microbenchmark results represent the one from \micronn-WarmCache scenario unless otherwise specified.
\begin{figure}[!h]
\centering
\begin{subfigure}{0.45\textwidth}
    \includegraphics[width=\textwidth]{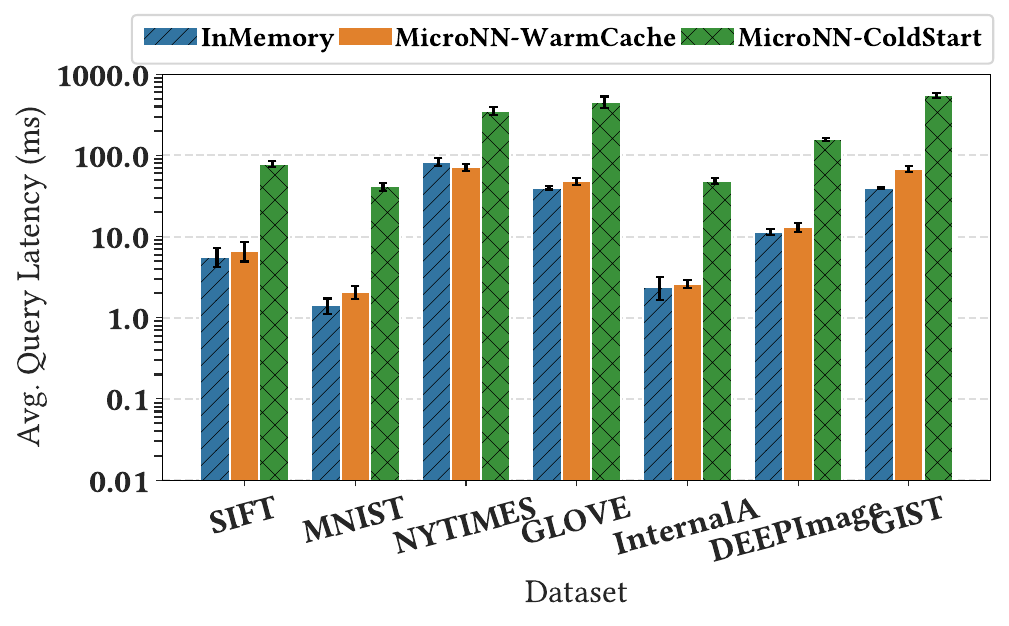}
    \caption{Large DUT}
    \label{fig:latency-large}
\end{subfigure}
\hfill
\begin{subfigure}{0.45\textwidth}
    \includegraphics[width=\textwidth]{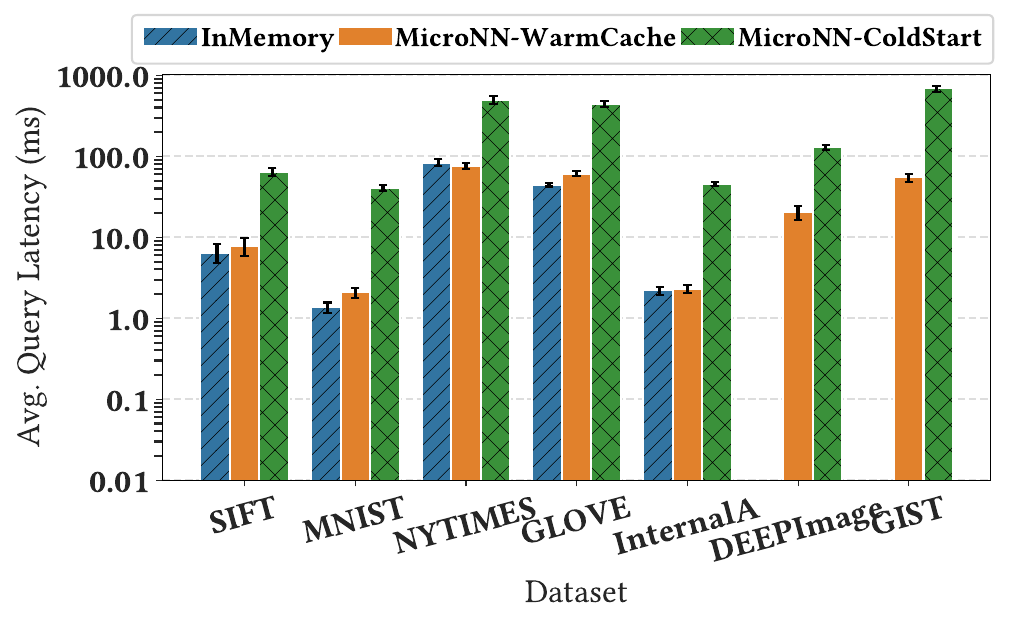}
    \caption{Small DUT}
    \label{fig:latency-small}
\end{subfigure}
\caption{Query latency for 90\% recall@100.}
\label{fig:e2e-latency}
\end{figure}

\subsection{End-to-end Performance}
We first demonstrate \micronn's capability to perform ANN search using a substantially small fraction of memory compared to a completely memory resident ANN system while exhibiting competitive query latency. We use all the datasets listed in Table~\ref{tab:dataset}. Note that the GIST and DEEPImage datasets could not be evaluated for the InMemory scenario on the Small DUT due to the device's physical memory limitations. 
This shows that \micronn is able to effectively utilize the disk and the memory to both build the ANN index and use the index for ANN search, even for the largest datasets. 

\subsubsection{Query Latency \& Memory Usage}
\begin{figure}[t]
\centering
\begin{subfigure}{0.41\textwidth}
    \includegraphics[width=\textwidth]{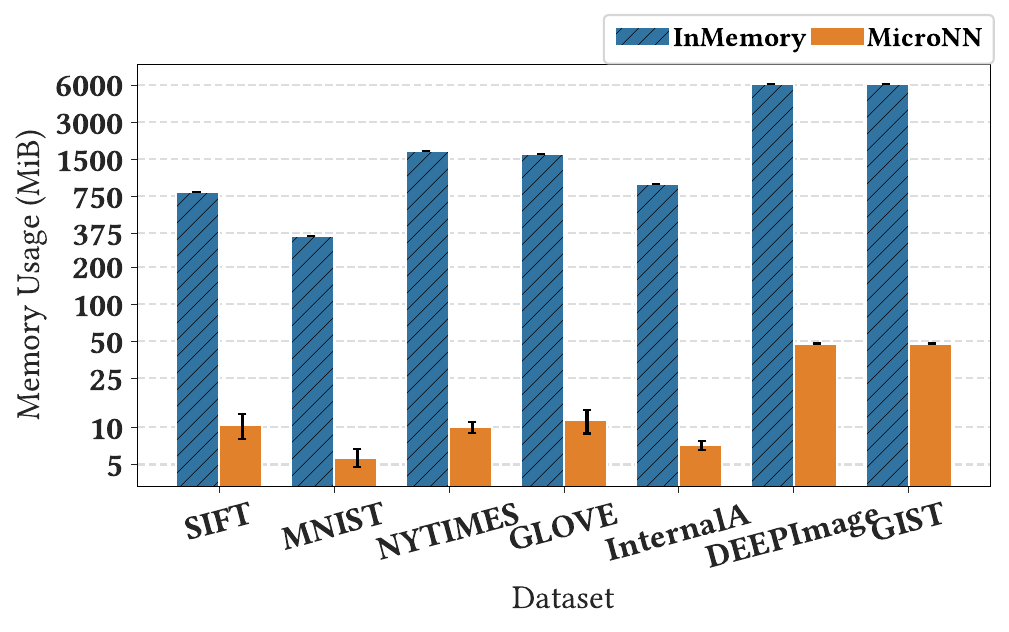}
    \caption{Large DUT}
    \label{fig:memory-large}
\end{subfigure}
\hfill
\begin{subfigure}{0.41\textwidth}
    \includegraphics[width=\textwidth]{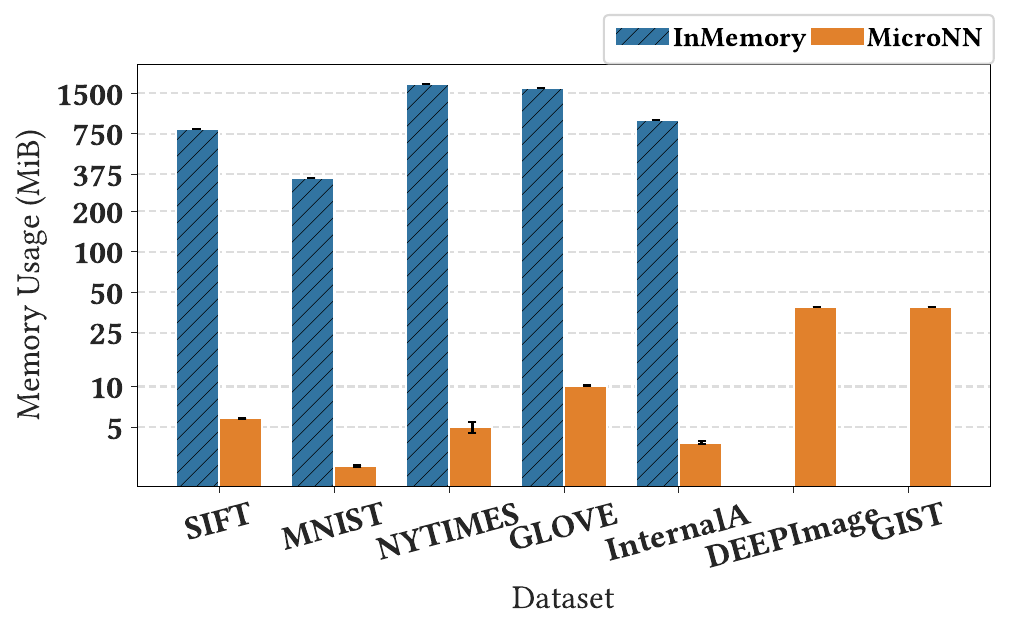}
    \caption{Small DUT}
    \label{fig:memory-small}
\end{subfigure}
\caption{Memory usage during query processing}
\label{fig:e2e-memory}
\end{figure}

Figure~\ref{fig:e2e-latency} shows the mean ANN search latency at 90\% recall@100 for InMemory, \micronn-ColdStart, and \micronn-WarmCache scenarios on a Large DUT. As expected, \micronn-ColdStart's latency in all datasets is an order of magnitude higher than the rest due to the cold centroid and database caches. However, as the database cache warms up, and the centroids get cached in memory, \micronn-WarmCache provides ANN search latency comparable to InMemory, while using two orders of magnitude less (as shown in Figure~\ref{fig:e2e-memory}).

\subsubsection{Index Construction Time \& Memory Usage}
\begin{figure}[t]
    \centering
    \begin{subfigure}{0.41\textwidth}
        \includegraphics[width=\textwidth]{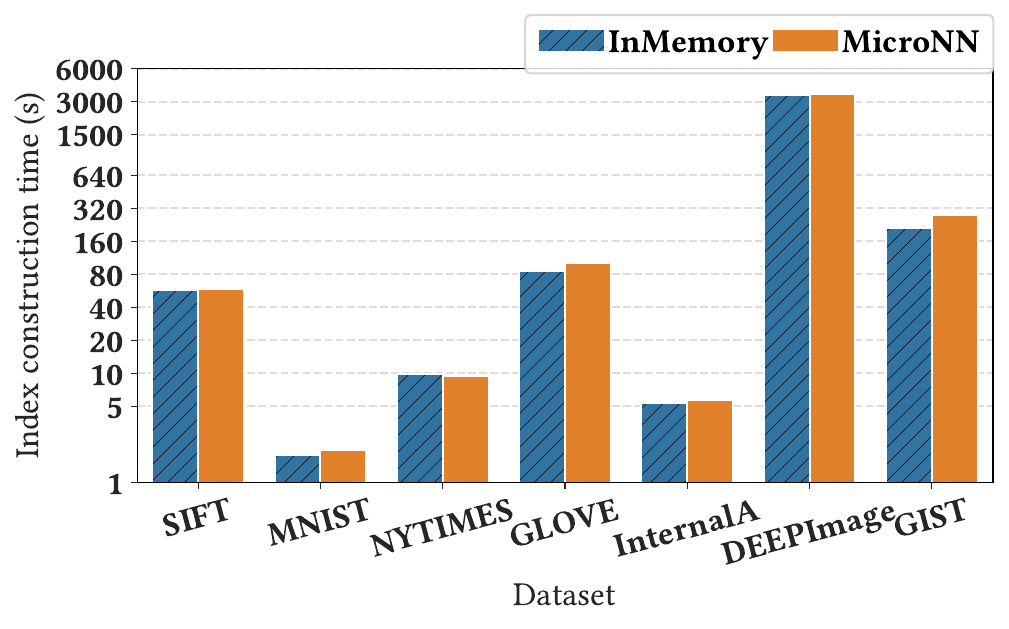}
        \caption{Index construction time}
        \label{fig:index-construction-time}
    \end{subfigure}
    \hfill
    \begin{subfigure}{0.41\textwidth}
        \includegraphics[width=\textwidth]{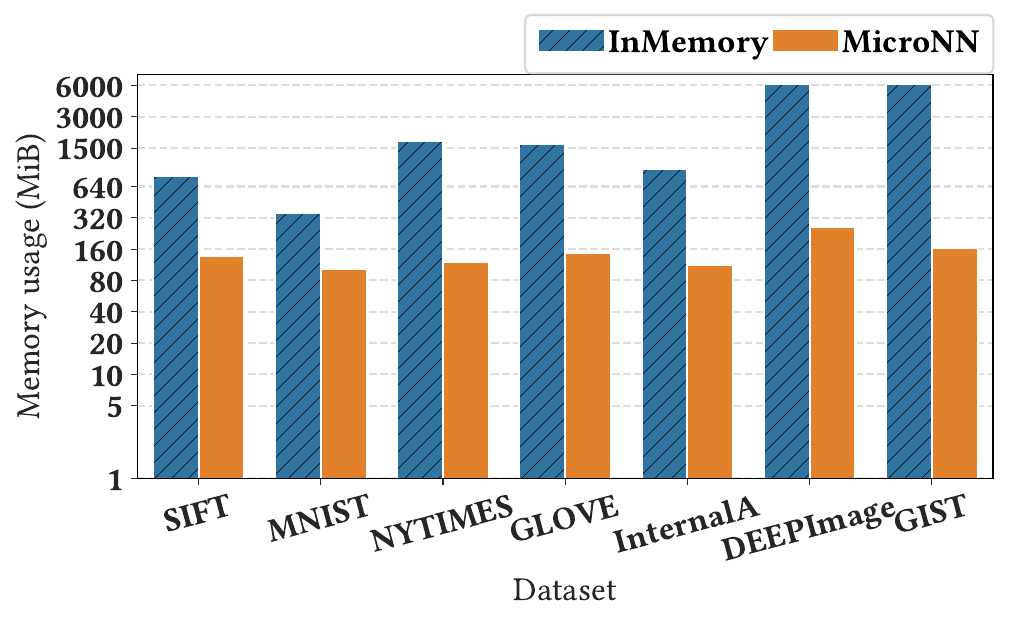}
        \caption{Memory usage during index construction}
        \label{fig:index-construction-memory}
    \end{subfigure}
    \caption{\micronn's index construction performance}
\label{fig:index-construction}
\end{figure}

The mini-batch k-means algorithm used in \micronn brings small batches of the embedding vectors from disk to memory while training the disk resident quantizer for the IVF index. In contrast, the InMemory approach needs to buffer all vectors in memory and thus has a significantly larger memory footprint as shown in Figure~\ref{fig:index-construction-memory}.
In addition, Figure~\ref{fig:index-construction-time} compares the index construction time of both approaches. We observe that memory and disk I/O has a negligible impact in index construction time as it is a compute intensive operation. These results highlight the benefits of a disk resident index construction mechanism for on-device deployments, where the index can be constructed using a fraction of memory without storing all vectors in memory.

\subsection{Microbenchmarks}

\subsubsection{Effectiveness of hybrid query optimizer}
\label{sec:experiements-optimizer}

We evaluate our query optimizer on the Big-ANN Filtered Search dataset~\cite{simhadri2024results}, which contains 10M CLIP~\cite{radford2021learning} embeddings of Flickr images. Each embedding is associated with a bag of tags. Each query for the benchmark contains a query embedding and a set of query tags, all of which must be associated with embeddings returned by the search.

We encode the tags as a whitespace separated string, then store the string as a string column in the Attributes table (from Figure~\ref{fig:example-schema}). An inverted index is built on this column, such that each tag is represented as a token. The attribute filters for each query is then a conjunction of \texttt{MATCH} filters on the string column, ensuring that the results must contain the query tags. Because the column is of string type, we use the string selectivity estimation method outlined in Section~\ref{sec:queryoptimizer} to estimate selectivity for hybrid query plan optimization. We set $n$ to 40 and use an average partition size of 500 for the IVF index. 

\begin{figure}
    \centering
    \begin{subfigure}{0.4\textwidth}
        \includegraphics[width=\textwidth]{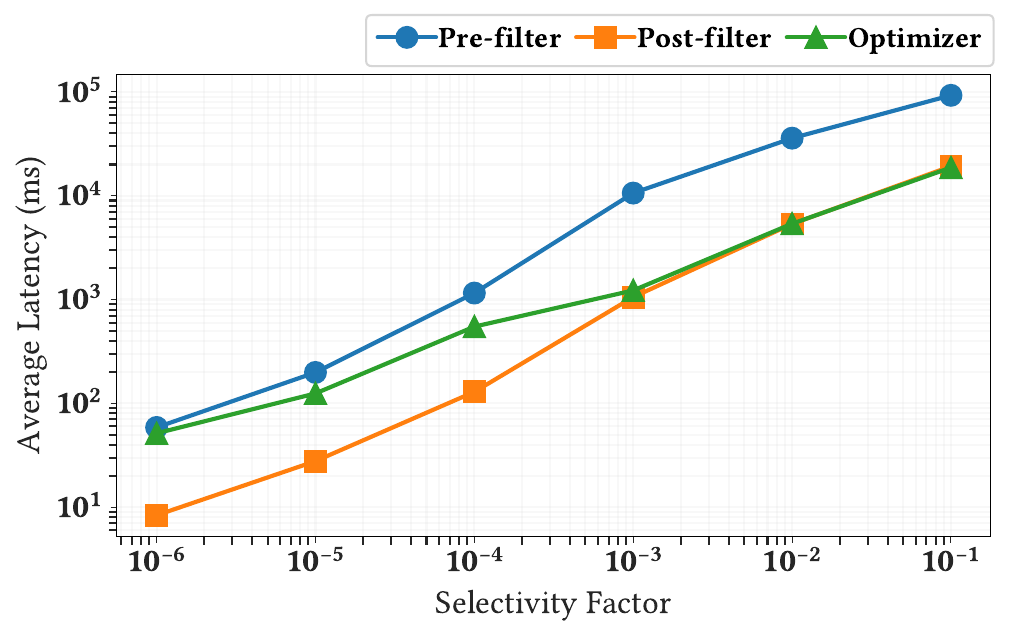}
        \caption{Latency vs. selectivity factor}
        \label{fig:selectivity-vs-latency}
    \end{subfigure}
    \hfill
    \begin{subfigure}{0.4\textwidth}
        \includegraphics[width=\textwidth]{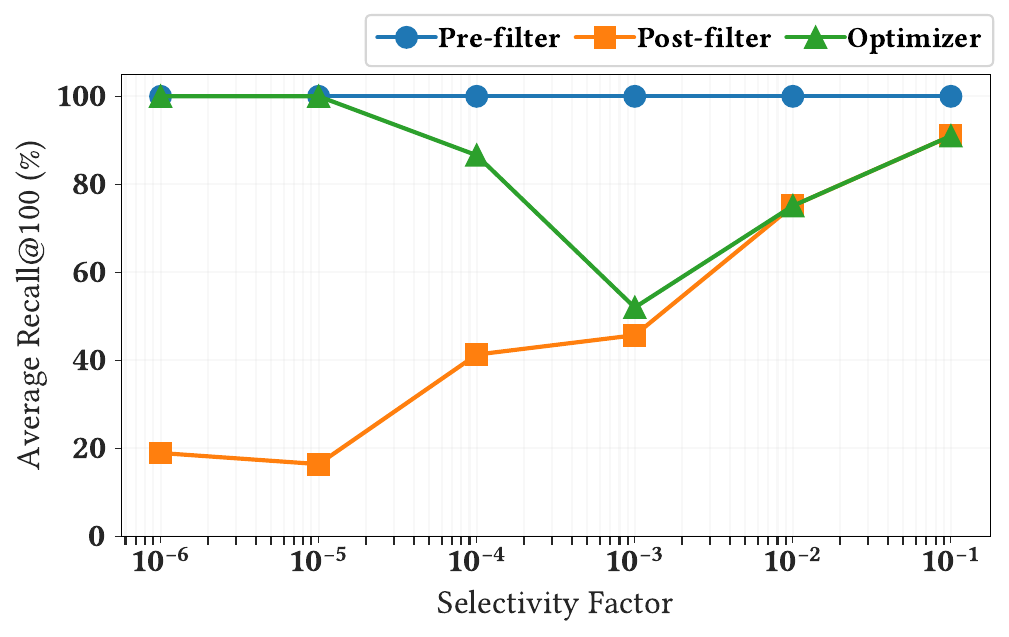}
        \caption{Recall@100 vs. selectivity factor. The client-defined optimizer threshold is set to bound latency to Pre-filter (before the threshold) and post-filter (after the threshold). Recall can be increased at the cost of additional latency by adjusting the threshold and the number of partition probes.}
        \label{fig:selectivity-vs-recall}
    \end{subfigure}
    \caption{Effectiveness of Hybrid Query Optimizer}
\label{fig:query-optimizer}
\end{figure}

From the provided queries, we measure the true predicate selectivity factor of each bag of query tags by executing the filters of the query and counting the number of rows in the result. We then bin the queries by their predicate selectivity factor order of magnitude, then sample 10 queries from each order of magnitude bin. We then execute these queries using the pre-filtering, post-filtering, and query optimizer based approaches, computing the average latencies and recall@100 within each bin. The results of this benchmark is presented in Figure \ref{fig:query-optimizer}.

Post-filtering is an order of magnitude faster than pre-filtering. We see that both methods get slower as they need to consider more vectors. However, post-filtering remains consistently faster relative to pre-filtering, as it is computing vector distances over a much smaller fraction of the total index. This can happen either because the predicate qualifies more vectors for pre-filtering than the post-filtering approach considers from the number of partitions scanned, or because many of the vectors scanned by post-filtering are filtered out by the attribute constraint and therefore no distance calculations are done on those vectors. 

While post-filtering is an order of magnitude faster, it suffers from very low recall for highly selective queries. This is because the IVF partitions, after applying the predicate filters, have significantly fewer vectors than the index expects to scan to achieve a reasonable recall. Because it scans and computes distances over fewer vectors, it is significantly faster than the pre-filtering strategy, however only few of the vectors scanned make their way into the top-K.

In contrast, pre-filtering has reasonably low latencies for highly selective queries while maintaining 100\% recall. As the number of vectors satisfying the filter increases, the latency of pre-filtering increases proportionally. At this point the recall of post-filtering becomes competitive at much lower latencies.

The query optimizer is able to effectively navigate this latency and recall trade-off. It achieves 100\% recall for highly selective queries, while being able to switch to the post-filtering strategy for less selective queries. This capability to switch to the appropriate plan leads to faster latencies than a pre-filtering only strategy, and higher recall than a post-filtering only strategy.

\subsubsection{Impact of memory usage on Index quality}
We evaluate the impact of using mini-batch k-means algorithm for training the IVF index's quantizer as opposed to using a k-means algorithm that considers all training vectors at once when training the quantizer. For this evaluation, we use the InternalA dataset and we vary the batch size used during training as the percentage of the dataset size and show how recall and memory usage during index construction are impacted as batch size changes. For recall computation, we identify the $n$ parameter (\ie the number of clusters to probe) to achieve 90\% recall on the index trained using the smallest batch size and use that $n$ throughout to ensure we perform roughly the same number of vector similarity computations.

\begin{figure}
\centering
\begin{subfigure}{0.4\textwidth}
    \includegraphics[width=\textwidth]{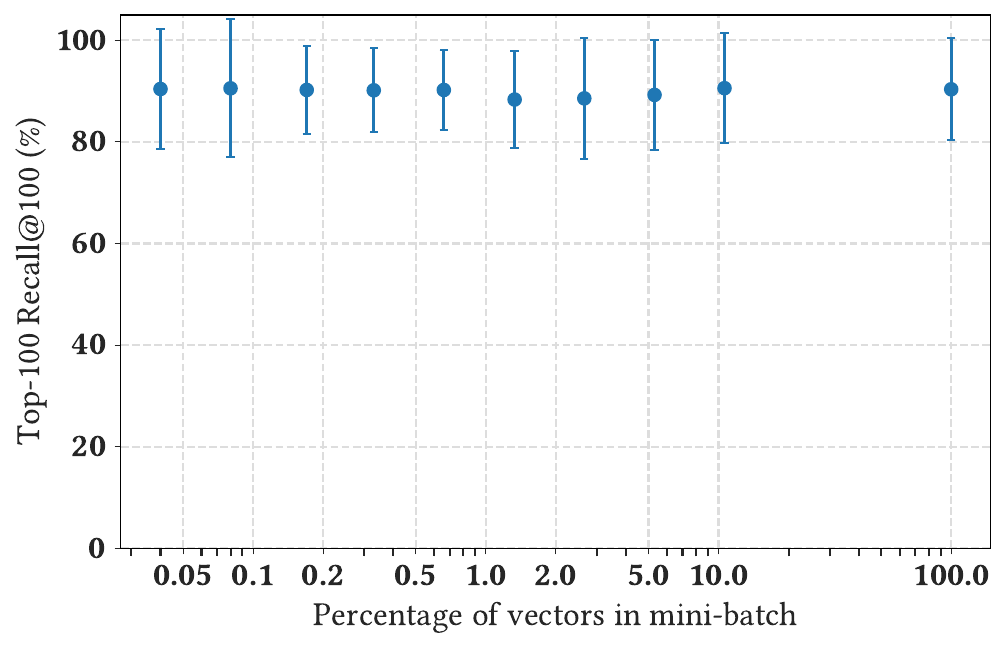}
    \caption{Recall of top-100 search}
    \label{fig:minibatch-impact-recall}
\end{subfigure}
\hfill
\begin{subfigure}{0.4\textwidth}
    \includegraphics[width=\textwidth]{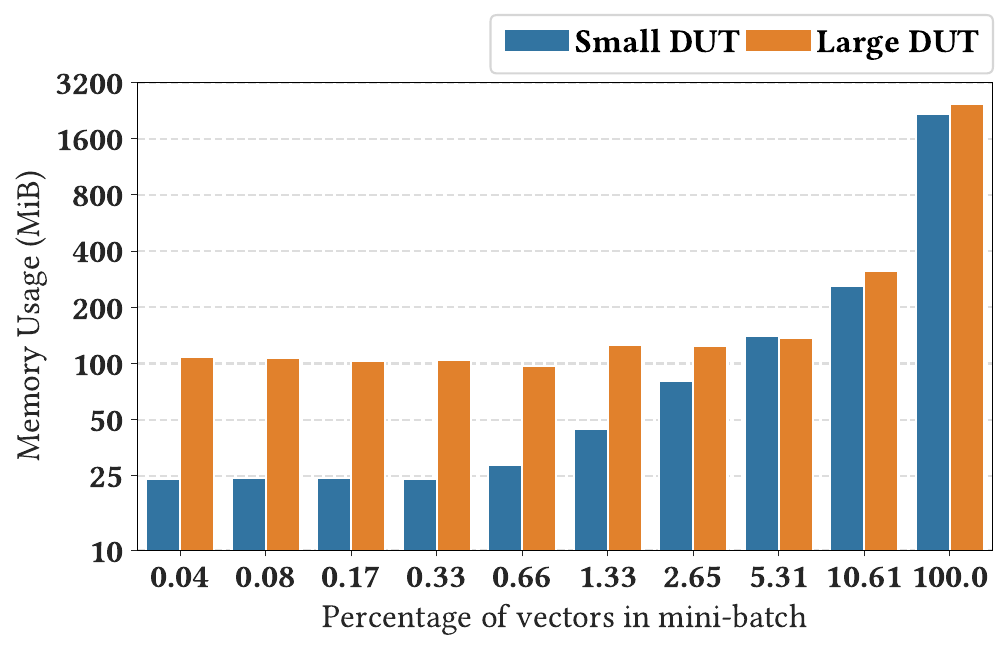}
    \caption{Memory Usage}
    \label{fig:minibatch-impact-memory}
\end{subfigure}
\caption{Impact of mini-batch k-means batch size on recall and memory usage}
\label{fig:minibatch-impact}
\end{figure}

As Figure~\ref{fig:minibatch-impact-recall} shows, there is little to no impact on recall as we vary the batch size from 0.04\% to all the way up-to 100\% of the training vectors, which resembles a regular k-means algorithm. Using a batch size of 0.04\% of the vectors still maintains a 90\% recall while using only 25 MiB of memory on the Small and 100 MiB of memory on the Large DUT, respectively. The difference in memory usage on the two device types is due to how they were configured for virtual memory page allocation as well as the SQLite page cache pool configuration. In contrast, a regular k-means algorithm (represented by 100\% mini-batch size) would use more than 1.6 GiB of memory for creating a clustering with similar quality.

\begin{figure}
    \centering
    \begin{subfigure}{0.42\textwidth}
        \includegraphics[width=\textwidth]{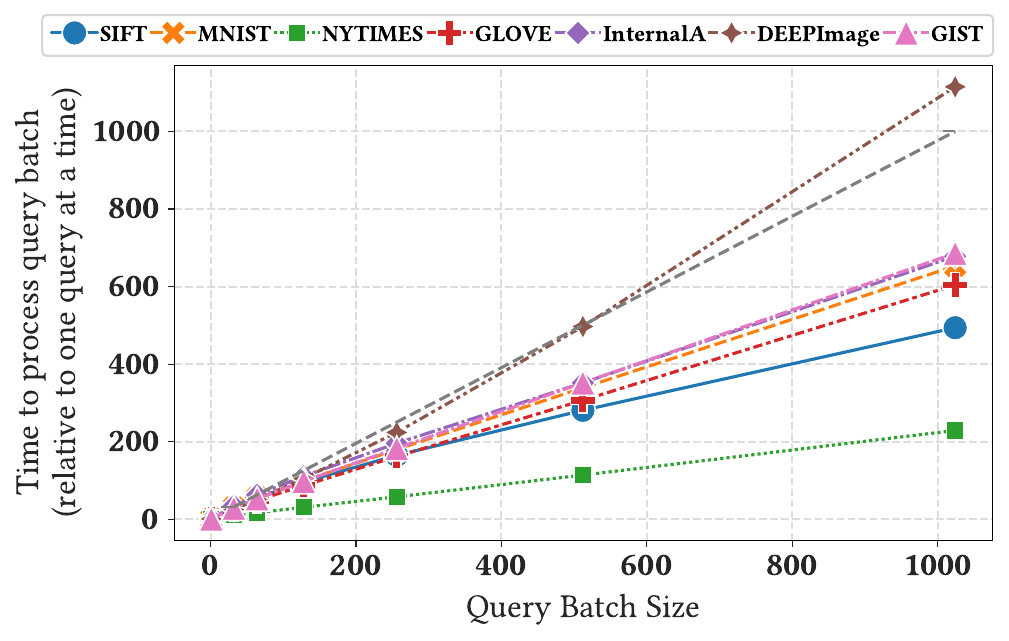}
        \caption{Batch processing time relative to sequential query processing}\label{fig:batch-completion-time}
    \end{subfigure}
    \begin{subfigure}{0.42\textwidth}
        \includegraphics[width=\textwidth]{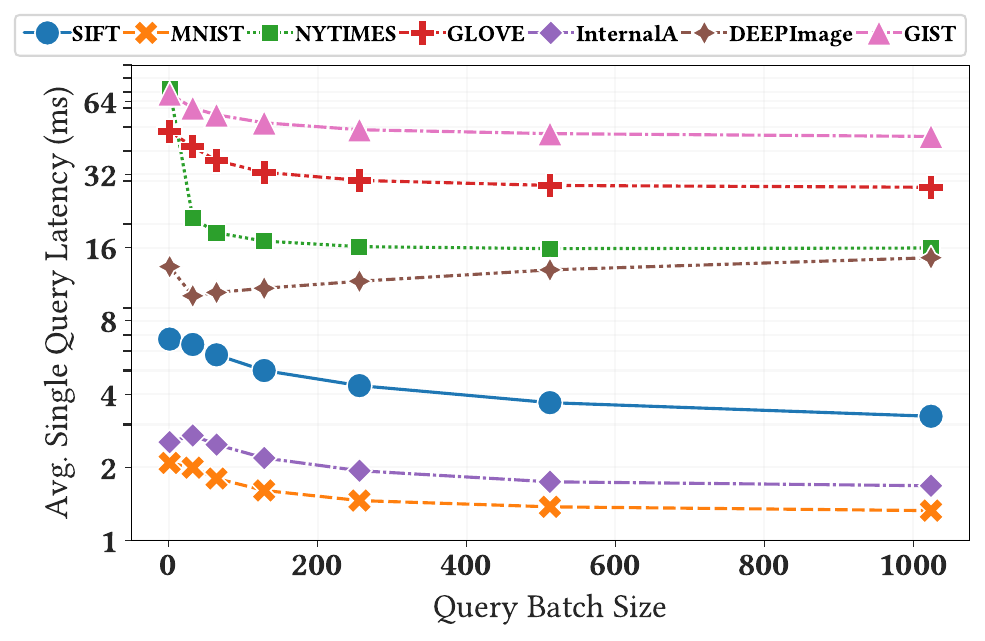}
        \caption{Amortized Single Query Latency}\label{fig:batch-amortized}
    \end{subfigure}
    \caption{Impact of Multi-Query Optimization}
    \label{fig:batch-query-result}
\end{figure}

\begin{figure}[!tbh]
    \centering
    \begin{subfigure}{0.235\textwidth}
        \includegraphics[width=\textwidth]{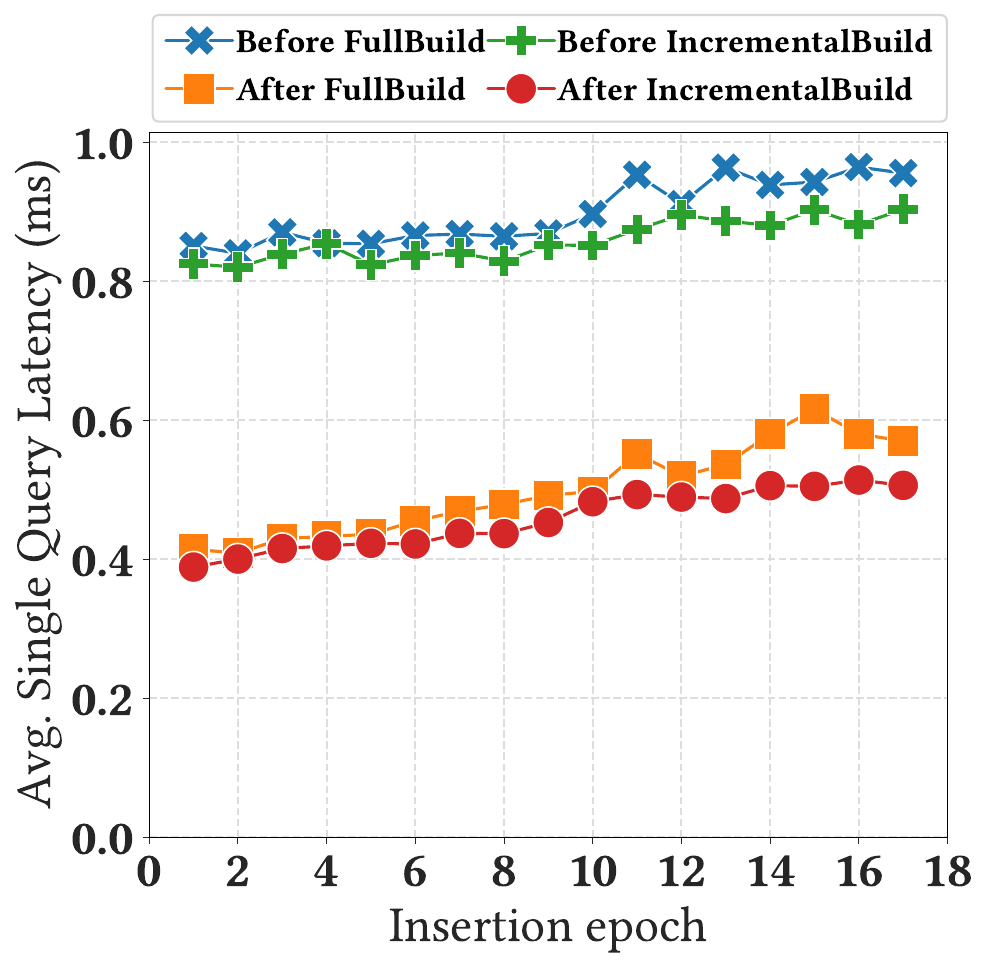}
        \caption{Avg. Single Query Latency}\label{fig:rebuild-query-latency}
    \end{subfigure}
    \begin{subfigure}{0.235\textwidth}
        \includegraphics[width=\textwidth]{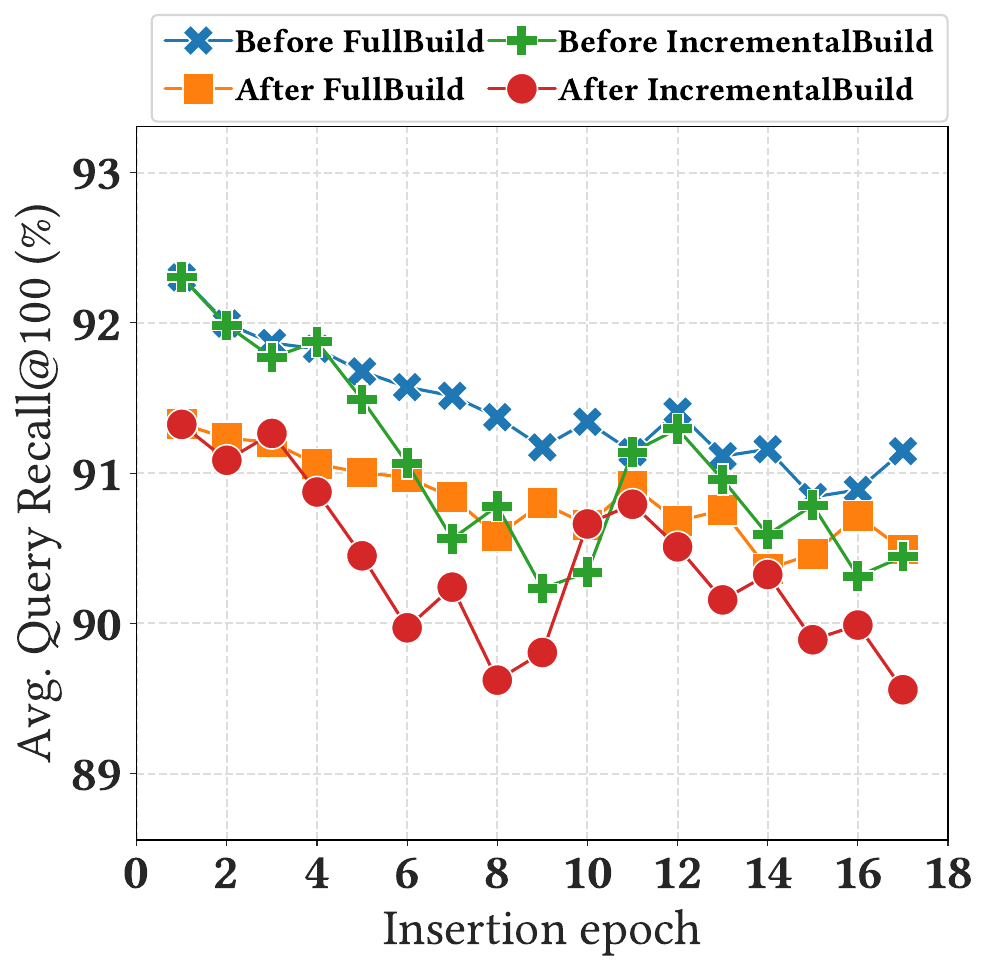}
        \caption{Recall of top-100 search}\label{fig:rebuild-recall}
    \end{subfigure}
    \begin{subfigure}{0.235\textwidth}
        \includegraphics[width=\textwidth]{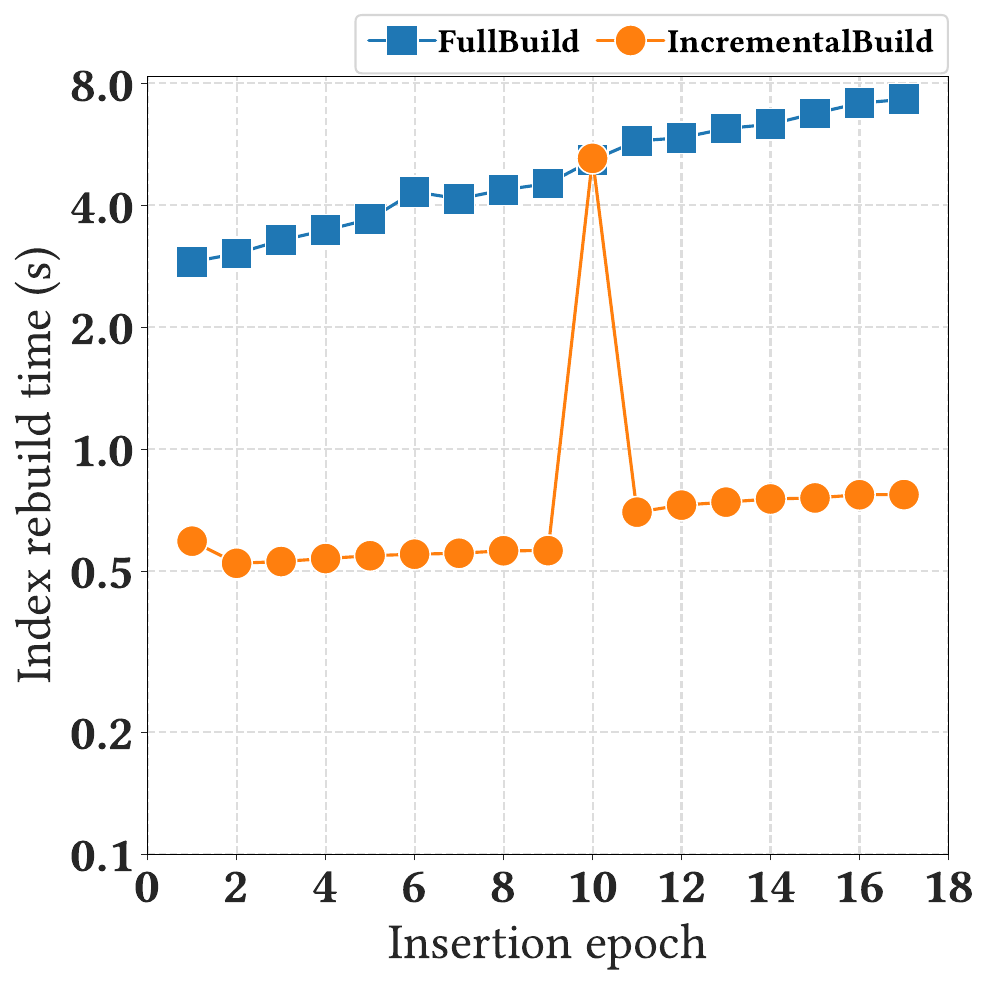}
        \caption{Index Build Time}\label{fig:rebuild-build-time}
    \end{subfigure}
    \begin{subfigure}{0.235\textwidth}
        \includegraphics[width=\textwidth]{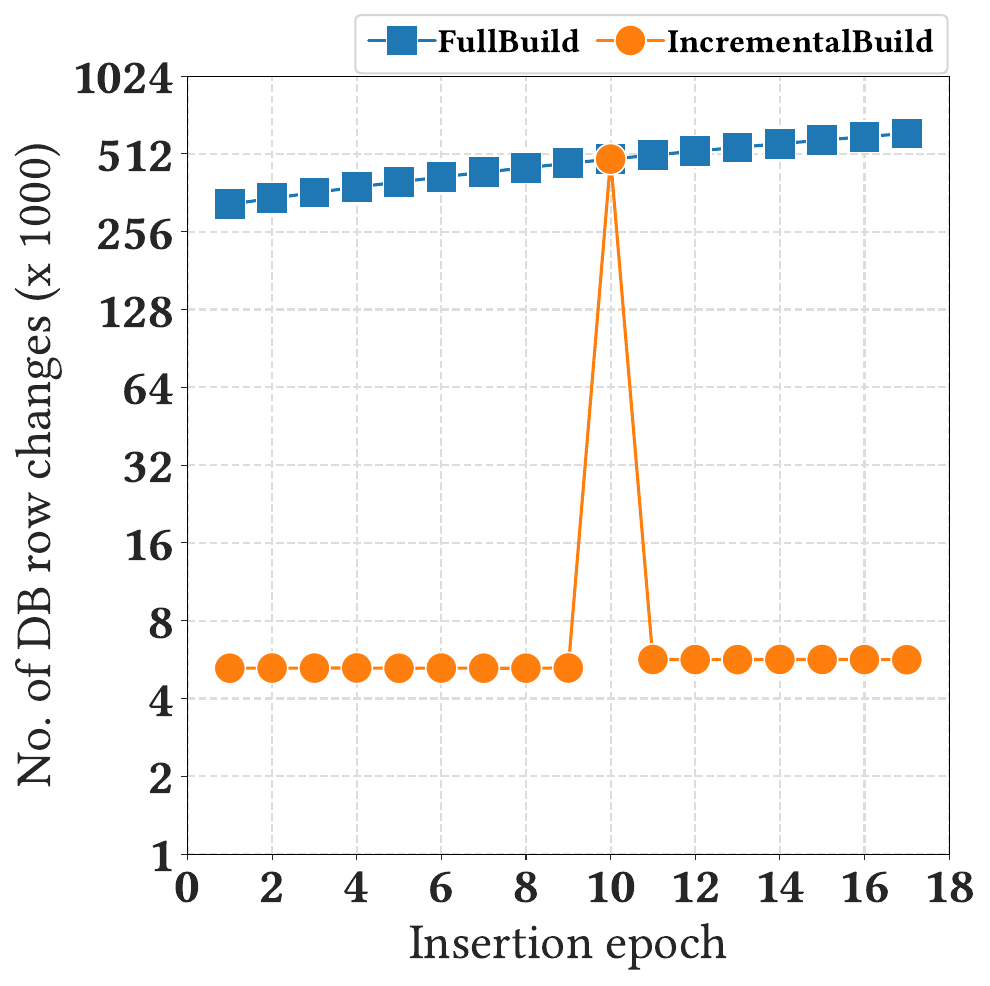}
        \caption{Number of Database Operations}\label{fig:rebuild-ops}
    \end{subfigure}
    \caption{Comparison between full and incremental index rebuild approaches on InternalA dataset}
    \label{fig:incremental-indexing}
\end{figure}

\subsubsection{Batch Query Execution}
We employ multi-query optimization techniques from~\cite{mohoney_high-throughput_2023} to batch execute multiple queries that need to compute similarity with vectors within the same IVF index partition. As such, \micronn can eliminate redundant scanning of IVF partitions and amortize data movement cost over a batch of queries. In addition, it takes advantage of vector instruction sets available on most modern CPUs for computing the similarities between a batch of query vectors and the vectors within an IVF partition. Figure~\ref{fig:batch-completion-time} shows the impact of our multi-query optimization on query latency for each of the datasets in Table~\ref{tab:dataset}. If queries were processed one after another, then theocratically the total time to process a batch would scale linearly as the batch size (shown by the dashed line in Figure~\ref{fig:batch-completion-time}). However, with the multi-query optimization in place, the total time for processing a query batch consistently less than processing each query in the batch one at a time. For instance, on InternalA, a batch size of 1024 is processed within 1.7 seconds achieving 90\% recall@100. Which results in an average latency of 1.7 ms per query, which is 33\% less compared to running queries one at a time. 

We observe that the gain diminishes as the matrix of query batch and cluster centroids grows larger. For instance, for the DEEPImage dataset with $\approx$100k centroids, the overhead of large matrix multiplication for a batch of 1024 outweighs the gains. This is an example of a scenario (mentioned earlier in Section~\ref{sec:physical-storage}) where additional indexing over the centroids would reduce the overhead of centroid scan, which is beyond the scope of this paper. For the other benchmarks, we see a larger batch size significantly improves completion time of analytical workloads.

\subsubsection{Index Updates}
We emulate the behavior of a growing vector collection by bootstrapping the IVF index using 50\% of the InternalA dataset, followed by inserting 3\% of the remaining dataset vectors into the index at each epoch. We compute recall at each epoch and use a query batch of size 128. We compare against an ideal scenario in which the index is fully rebuilt after each epoch. For our incremental update approach, we set a threshold of 50\% increase in average partition size to trigger a full index rebuild. Due to increase in average partition size, the number of vectors scanned increases if $n$ is fixed, which also increases the recall. In our experiments, we keep updating $n$ to keep the target number of vectors scanned same throughout. Results are presented in Figure~\ref{fig:incremental-indexing}.

Since $n$ is adjusted throughout the experiment, the average single query latency remains comparable with the full rebuild approach (Figure~\ref{fig:rebuild-query-latency}). For both approaches, the query latency before index rebuild was high due to exhaustive scanning of the delta-store during query processing. The impact in this case is more on recall of top-100 search (Figure~\ref{fig:rebuild-recall}). With both approaches it is expected that query recall will drop after the delta-store is flushed since we are now performing ANN search over the vectors previously in the delta-store. As the index is updated, incremental rebuild mechanism keeps deviating from the ideal recall of a full rebuild. However, the deviation remains small and is corrected as soon as a full rebuild is triggered. This small loss in recall is compensated by the faster index build time (Figure~\ref{fig:rebuild-build-time}) and the significantly smaller I/O footprint (<2\% of full rebuild) of the incremental update approach (Figure~\ref{fig:rebuild-ops}). We see the incremental rebuild approach has comparable rebuild cost to full rebuild at the 10$^{\textrm{th}}$ epoch when the full rebuild criteria is triggered.

%% file: sections/conclusion.tex
\section{Conclusion}\label{sec:conclusion}
We have presented \micronn, an on-device embedded vector search engine that supports scalable similarity search under strict memory constraints. \micronn uses a disk-resident index and query processing algorithm, and includes support for updates, hybrid search queries with structured attribute filters, and optimizations for batch query processing all over the same index instance. 

We have shown that the overhead of a disk-resident vector index can be mitigated by mapping vector partitioning schemes onto relational database support for clustered indices, providing competitive recall and query latencies with orders of magnitude less memory usage. We have also demonstrated adaptations to traditional IVF indexing approaches that allows constructing and maintaining the index with minimal memory and disk I/O. We showed how a simple query optimizer can efficiently navigate pre- and post-filtering query plans, and how multi-query optimization techniques can be leveraged to accelerate batch query processing.

\todo{restore for camera ready}
\section{Acknowledgments}
 We would like to acknowledge the many contributions of our collaborators at Apple that helped make this project a success, including Gautam Sakleshpur Muralidhar, Sanjay Mohan, Xun Shi, Hongting Wang, Pranav Prashant Thombre, Riddhi Zunjarrao, Christine O'Mara, Sayantan Mahinder, Chiraag Sumanth, and JP Lacerda.
\todo{names}

%% file: main.bbl

\begin{thebibliography}{44}


\ifx \showCODEN    \undefined \def \showCODEN     #1{\unskip}     \fi
\ifx \showDOI      \undefined \def \showDOI       #1{#1}\fi
\ifx \showISBNx    \undefined \def \showISBNx     #1{\unskip}     \fi
\ifx \showISBNxiii \undefined \def \showISBNxiii  #1{\unskip}     \fi
\ifx \showISSN     \undefined \def \showISSN      #1{\unskip}     \fi
\ifx \showLCCN     \undefined \def \showLCCN      #1{\unskip}     \fi
\ifx \shownote     \undefined \def \shownote      #1{#1}          \fi
\ifx \showarticletitle \undefined \def \showarticletitle #1{#1}   \fi
\ifx \showURL      \undefined \def \showURL       {\relax}        \fi
\providecommand\bibfield[2]{#2}
\providecommand\bibinfo[2]{#2}
\providecommand\natexlab[1]{#1}
\providecommand\showeprint[2][]{arXiv:#2}

\bibitem[Arandjelovic and Zisserman(2013)]%
        {arandjelovic2013all}
\bibfield{author}{\bibinfo{person}{Relja Arandjelovic} {and} \bibinfo{person}{Andrew Zisserman}.} \bibinfo{year}{2013}\natexlab{}.
\newblock \showarticletitle{All about VLAD}. In \bibinfo{booktitle}{\emph{Proceedings of the IEEE conference on Computer Vision and Pattern Recognition}}. \bibinfo{publisher}{IEEE}, \bibinfo{pages}{1578--1585}.
\newblock


\bibitem[Arora et~al\mbox{.}(2018)]%
        {10.14778/3204028.3204034}
\bibfield{author}{\bibinfo{person}{Akhil Arora}, \bibinfo{person}{Sakshi Sinha}, \bibinfo{person}{Piyush Kumar}, {and} \bibinfo{person}{Arnab Bhattacharya}.} \bibinfo{year}{2018}\natexlab{}.
\newblock \showarticletitle{HD-index: pushing the scalability-accuracy boundary for approximate kNN search in high-dimensional spaces}.
\newblock \bibinfo{journal}{\emph{Proc. VLDB Endow.}} \bibinfo{volume}{11}, \bibinfo{number}{8} (\bibinfo{date}{April} \bibinfo{year}{2018}), \bibinfo{pages}{906–919}.
\newblock
\showISSN{2150-8097}
\urldef\tempurl%
\url{https://doi.org/10.14778/3204028.3204034}
\showDOI{\tempurl}


\bibitem[Babenko and Lempitsky(2015)]%
        {babenko_inverted_2015}
\bibfield{author}{\bibinfo{person}{Artem Babenko} {and} \bibinfo{person}{Victor Lempitsky}.} \bibinfo{year}{2015}\natexlab{}.
\newblock \showarticletitle{The {Inverted} {Multi}-{Index}}.
\newblock \bibinfo{journal}{\emph{IEEE Transactions on Pattern Analysis and Machine Intelligence}} \bibinfo{volume}{37}, \bibinfo{number}{6} (\bibinfo{date}{June} \bibinfo{year}{2015}), \bibinfo{pages}{1247--1260}.
\newblock
\showISSN{0162-8828, 2160-9292}
\urldef\tempurl%
\url{https://doi.org/10.1109/TPAMI.2014.2361319}
\showDOI{\tempurl}


\bibitem[Baranchuk et~al\mbox{.}(2018)]%
        {ferrari_revisiting_2018}
\bibfield{author}{\bibinfo{person}{Dmitry Baranchuk}, \bibinfo{person}{Artem Babenko}, {and} \bibinfo{person}{Yury Malkov}.} \bibinfo{year}{2018}\natexlab{}.
\newblock \showarticletitle{Revisiting the {Inverted} {Indices} for {Billion}-{Scale} {Approximate} {Nearest} {Neighbors}}.
\newblock In \bibinfo{booktitle}{\emph{Computer {Vision} – {ECCV} 2018}}, \bibfield{editor}{\bibinfo{person}{Vittorio Ferrari}, \bibinfo{person}{Martial Hebert}, \bibinfo{person}{Cristian Sminchisescu}, {and} \bibinfo{person}{Yair Weiss}} (Eds.). Vol.~\bibinfo{volume}{11216}. \bibinfo{publisher}{Springer International Publishing}, \bibinfo{address}{Cham}, \bibinfo{pages}{209--224}.
\newblock
\showISBNx{978-3-030-01257-1 978-3-030-01258-8}
\urldef\tempurl%
\url{https://doi.org/10.1007/978-3-030-01258-8_13}
\showDOI{\tempurl}
\newblock
\shownote{Series Title: Lecture Notes in Computer Science}.


\bibitem[Bernhardsson({[n.\,d.]})]%
        {annoy}
\bibfield{author}{\bibinfo{person}{Erik Bernhardsson}.} \bibinfo{year}{[n.\,d.]}\natexlab{}.
\newblock \bibinfo{title}{spotify/annoy: Approximate Nearest Neighbors in C++/Python.}
\newblock
\newblock
\urldef\tempurl%
\url{https://github.com/spotify/annoy}
\showURL{%
\tempurl}


\bibitem[Chen et~al\mbox{.}(2021)]%
        {chen_spann_nodate}
\bibfield{author}{\bibinfo{person}{Qi Chen}, \bibinfo{person}{Bing Zhao}, \bibinfo{person}{Haidong Wang}, \bibinfo{person}{Mingqin Li}, \bibinfo{person}{Chuanjie Liu}, \bibinfo{person}{Zengzhong Li}, \bibinfo{person}{Mao Yang}, {and} \bibinfo{person}{Jingdong Wang}.} \bibinfo{year}{2021}\natexlab{}.
\newblock \showarticletitle{SPANN: Highly-efficient Billion-scale Approximate Nearest Neighborhood Search}. In \bibinfo{booktitle}{\emph{Advances in Neural Information Processing Systems}}, \bibfield{editor}{\bibinfo{person}{M.~Ranzato}, \bibinfo{person}{A.~Beygelzimer}, \bibinfo{person}{Y.~Dauphin}, \bibinfo{person}{P.S. Liang}, {and} \bibinfo{person}{J.~Wortman Vaughan}} (Eds.), Vol.~\bibinfo{volume}{34}. \bibinfo{publisher}{Curran Associates, Inc.}, \bibinfo{pages}{5199--5212}.
\newblock
\urldef\tempurl%
\url{https://proceedings.neurips.cc/paper_files/paper/2021/file/299dc35e747eb77177d9cea10a802da2-Paper.pdf}
\showURL{%
\tempurl}


\bibitem[Du et~al\mbox{.}(2022)]%
        {10.1145/3534678.3539071}
\bibfield{author}{\bibinfo{person}{Ming Du}, \bibinfo{person}{Arnau Ramisa}, \bibinfo{person}{Amit~Kumar K~C}, \bibinfo{person}{Sampath Chanda}, \bibinfo{person}{Mengjiao Wang}, \bibinfo{person}{Neelakandan Rajesh}, \bibinfo{person}{Shasha Li}, \bibinfo{person}{Yingchuan Hu}, \bibinfo{person}{Tao Zhou}, \bibinfo{person}{Nagashri Lakshminarayana}, \bibinfo{person}{Son Tran}, {and} \bibinfo{person}{Doug Gray}.} \bibinfo{year}{2022}\natexlab{}.
\newblock \showarticletitle{Amazon Shop the Look: A Visual Search System for Fashion and Home}. In \bibinfo{booktitle}{\emph{Proceedings of the 28th ACM SIGKDD Conference on Knowledge Discovery and Data Mining}} (Washington DC, USA) \emph{(\bibinfo{series}{KDD '22})}. \bibinfo{publisher}{Association for Computing Machinery}, \bibinfo{address}{New York, NY, USA}, \bibinfo{pages}{2822–2830}.
\newblock
\showISBNx{9781450393850}
\urldef\tempurl%
\url{https://doi.org/10.1145/3534678.3539071}
\showDOI{\tempurl}


\bibitem[Friedman et~al\mbox{.}(1976)]%
        {kd-tree}
\bibfield{author}{\bibinfo{person}{Jerome~H Friedman}, \bibinfo{person}{Jon~Louis Bentley}, {and} \bibinfo{person}{Raphael~Ari Finkel}.} \bibinfo{year}{1976}\natexlab{}.
\newblock \showarticletitle{An algorithm for finding best matches in logarithmic time}.
\newblock \bibinfo{journal}{\emph{ACM Trans. Math. Software}} \bibinfo{volume}{3}, \bibinfo{number}{SLAC-PUB-1549-REV. 2} (\bibinfo{year}{1976}), \bibinfo{pages}{209--226}.
\newblock


\bibitem[Fu et~al\mbox{.}(2019)]%
        {fu12fast}
\bibfield{author}{\bibinfo{person}{Cong Fu}, \bibinfo{person}{Chao Xiang}, \bibinfo{person}{Changxu Wang}, {and} \bibinfo{person}{Deng Cai}.} \bibinfo{year}{2019}\natexlab{}.
\newblock \showarticletitle{Fast approximate nearest neighbor search with the navigating spreading-out graph}.
\newblock \bibinfo{journal}{\emph{Proc. VLDB Endow.}} \bibinfo{volume}{12}, \bibinfo{number}{5} (\bibinfo{date}{Jan.} \bibinfo{year}{2019}), \bibinfo{pages}{461–474}.
\newblock
\showISSN{2150-8097}
\urldef\tempurl%
\url{https://doi.org/10.14778/3303753.3303754}
\showDOI{\tempurl}


\bibitem[Gan et~al\mbox{.}(2023)]%
        {10.1145/3580305.3599782}
\bibfield{author}{\bibinfo{person}{Yukang Gan}, \bibinfo{person}{Yixiao Ge}, \bibinfo{person}{Chang Zhou}, \bibinfo{person}{Shupeng Su}, \bibinfo{person}{Zhouchuan Xu}, \bibinfo{person}{Xuyuan Xu}, \bibinfo{person}{Quanchao Hui}, \bibinfo{person}{Xiang Chen}, \bibinfo{person}{Yexin Wang}, {and} \bibinfo{person}{Ying Shan}.} \bibinfo{year}{2023}\natexlab{}.
\newblock \showarticletitle{Binary Embedding-based Retrieval at Tencent}. In \bibinfo{booktitle}{\emph{Proceedings of the 29th ACM SIGKDD Conference on Knowledge Discovery and Data Mining}} (Long Beach, CA, USA) \emph{(\bibinfo{series}{KDD '23})}. \bibinfo{publisher}{Association for Computing Machinery}, \bibinfo{address}{New York, NY, USA}, \bibinfo{pages}{4056–4067}.
\newblock
\showISBNx{9798400701030}
\urldef\tempurl%
\url{https://doi.org/10.1145/3580305.3599782}
\showDOI{\tempurl}


\bibitem[Gollapudi et~al\mbox{.}(2023)]%
        {gollapudi2023filtered}
\bibfield{author}{\bibinfo{person}{Siddharth Gollapudi}, \bibinfo{person}{Neel Karia}, \bibinfo{person}{Varun Sivashankar}, \bibinfo{person}{Ravishankar Krishnaswamy}, \bibinfo{person}{Nikit Begwani}, \bibinfo{person}{Swapnil Raz}, \bibinfo{person}{Yiyong Lin}, \bibinfo{person}{Yin Zhang}, \bibinfo{person}{Neelam Mahapatro}, \bibinfo{person}{Premkumar Srinivasan}, \bibinfo{person}{Amit Singh}, {and} \bibinfo{person}{Harsha~Vardhan Simhadri}.} \bibinfo{year}{2023}\natexlab{}.
\newblock \showarticletitle{Filtered-DiskANN: Graph Algorithms for Approximate Nearest Neighbor Search with Filters}. In \bibinfo{booktitle}{\emph{Proceedings of the ACM Web Conference 2023}} (Austin, TX, USA) \emph{(\bibinfo{series}{WWW '23})}. \bibinfo{publisher}{Association for Computing Machinery}, \bibinfo{address}{New York, NY, USA}, \bibinfo{pages}{3406–3416}.
\newblock
\showISBNx{9781450394161}
\urldef\tempurl%
\url{https://doi.org/10.1145/3543507.3583552}
\showDOI{\tempurl}


\bibitem[Grbovic and Cheng(2018)]%
        {grbovic2018real}
\bibfield{author}{\bibinfo{person}{Mihajlo Grbovic} {and} \bibinfo{person}{Haibin Cheng}.} \bibinfo{year}{2018}\natexlab{}.
\newblock \showarticletitle{Real-time personalization using embeddings for search ranking at airbnb}. In \bibinfo{booktitle}{\emph{Proceedings of the 24th ACM SIGKDD International Conference on Knowledge Discovery \& Data Mining}}. \bibinfo{pages}{311--320}.
\newblock


\bibitem[Guo et~al\mbox{.}(2020)]%
        {guo_accelerating_2020}
\bibfield{author}{\bibinfo{person}{Ruiqi Guo}, \bibinfo{person}{Philip Sun}, \bibinfo{person}{Erik Lindgren}, \bibinfo{person}{Quan Geng}, \bibinfo{person}{David Simcha}, \bibinfo{person}{Felix Chern}, {and} \bibinfo{person}{Sanjiv Kumar}.} \bibinfo{year}{2020}\natexlab{}.
\newblock \showarticletitle{Accelerating {Large}-{Scale} {Inference} with {Anisotropic} {Vector} {Quantization}}. In \bibinfo{booktitle}{\emph{Proceedings of the 37th {International} {Conference} on {Machine} {Learning}}}. \bibinfo{publisher}{PMLR}, \bibinfo{pages}{3887--3896}.
\newblock
\urldef\tempurl%
\url{https://proceedings.mlr.press/v119/guo20h.html}
\showURL{%
\tempurl}
\newblock
\shownote{ISSN: 2640-3498}.


\bibitem[Haldar et~al\mbox{.}(2019)]%
        {haldar2019applying}
\bibfield{author}{\bibinfo{person}{Malay Haldar}, \bibinfo{person}{Mustafa Abdool}, \bibinfo{person}{Prashant Ramanathan}, \bibinfo{person}{Tao Xu}, \bibinfo{person}{Shulin Yang}, \bibinfo{person}{Huizhong Duan}, \bibinfo{person}{Qing Zhang}, \bibinfo{person}{Nick Barrow-Williams}, \bibinfo{person}{Bradley~C Turnbull}, \bibinfo{person}{Brendan~M Collins}, {et~al\mbox{.}}} \bibinfo{year}{2019}\natexlab{}.
\newblock \showarticletitle{Applying deep learning to airbnb search}. In \bibinfo{booktitle}{\emph{Proceedings of the 25th ACM SIGKDD International Conference on Knowledge Discovery \& Data Mining}}. \bibinfo{pages}{1927--1935}.
\newblock


\bibitem[Hashemi et~al\mbox{.}(2021)]%
        {hashemi2021neural}
\bibfield{author}{\bibinfo{person}{Helia Hashemi}, \bibinfo{person}{Aasish Pappu}, \bibinfo{person}{Mi Tian}, \bibinfo{person}{Praveen Chandar}, \bibinfo{person}{Mounia Lalmas}, {and} \bibinfo{person}{Benjamin Carterette}.} \bibinfo{year}{2021}\natexlab{}.
\newblock \showarticletitle{Neural instant search for music and podcast}. In \bibinfo{booktitle}{\emph{Proceedings of the 27th ACM SIGKDD Conference on Knowledge Discovery \& Data Mining}}. \bibinfo{pages}{2984--2992}.
\newblock


\bibitem[Hossain et~al\mbox{.}(2019)]%
        {hossain2019comprehensive}
\bibfield{author}{\bibinfo{person}{MD~Zakir Hossain}, \bibinfo{person}{Ferdous Sohel}, \bibinfo{person}{Mohd~Fairuz Shiratuddin}, {and} \bibinfo{person}{Hamid Laga}.} \bibinfo{year}{2019}\natexlab{}.
\newblock \showarticletitle{A comprehensive survey of deep learning for image captioning}.
\newblock \bibinfo{journal}{\emph{ACM Computing Surveys (CsUR)}} \bibinfo{volume}{51}, \bibinfo{number}{6} (\bibinfo{year}{2019}), \bibinfo{pages}{1--36}.
\newblock


\bibitem[Jayaram~Subramanya et~al\mbox{.}(2019)]%
        {jayaram2019diskann}
\bibfield{author}{\bibinfo{person}{Suhas Jayaram~Subramanya}, \bibinfo{person}{Fnu Devvrit}, \bibinfo{person}{Harsha~Vardhan Simhadri}, \bibinfo{person}{Ravishankar Krishnawamy}, {and} \bibinfo{person}{Rohan Kadekodi}.} \bibinfo{year}{2019}\natexlab{}.
\newblock \showarticletitle{Diskann: Fast accurate billion-point nearest neighbor search on a single node}.
\newblock \bibinfo{journal}{\emph{Advances in Neural Information Processing Systems}}  \bibinfo{volume}{32} (\bibinfo{year}{2019}).
\newblock


\bibitem[Johnson et~al\mbox{.}(2019)]%
        {johnson2019billion}
\bibfield{author}{\bibinfo{person}{Jeff Johnson}, \bibinfo{person}{Matthijs Douze}, {and} \bibinfo{person}{Herv{\'e} J{\'e}gou}.} \bibinfo{year}{2019}\natexlab{}.
\newblock \showarticletitle{Billion-scale similarity search with GPUs}.
\newblock \bibinfo{journal}{\emph{IEEE Transactions on Big Data}} \bibinfo{volume}{7}, \bibinfo{number}{3} (\bibinfo{year}{2019}), \bibinfo{pages}{535--547}.
\newblock


\bibitem[Jégou et~al\mbox{.}(2011a)]%
        {jegou_product_2011}
\bibfield{author}{\bibinfo{person}{Herve Jégou}, \bibinfo{person}{Matthijs Douze}, {and} \bibinfo{person}{Cordelia Schmid}.} \bibinfo{year}{2011}\natexlab{a}.
\newblock \showarticletitle{Product {Quantization} for {Nearest} {Neighbor} {Search}}.
\newblock \bibinfo{journal}{\emph{IEEE Transactions on Pattern Analysis and Machine Intelligence}} \bibinfo{volume}{33}, \bibinfo{number}{1} (\bibinfo{date}{Jan.} \bibinfo{year}{2011}), \bibinfo{pages}{117--128}.
\newblock
\showISSN{1939-3539}
\urldef\tempurl%
\url{https://doi.org/10.1109/TPAMI.2010.57}
\showDOI{\tempurl}
\newblock
\shownote{Conference Name: IEEE Transactions on Pattern Analysis and Machine Intelligence}.


\bibitem[Jégou et~al\mbox{.}(2011b)]%
        {jégou2011searching}
\bibfield{author}{\bibinfo{person}{Hervé Jégou}, \bibinfo{person}{Romain Tavenard}, \bibinfo{person}{Matthijs Douze}, {and} \bibinfo{person}{Laurent Amsaleg}.} \bibinfo{year}{2011}\natexlab{b}.
\newblock \bibinfo{title}{Searching in one billion vectors: re-rank with source coding}.
\newblock
\newblock
\showeprint[arxiv]{1102.3828}~[cs.IR]


\bibitem[Liu et~al\mbox{.}(2017)]%
        {liu2017related}
\bibfield{author}{\bibinfo{person}{David~C Liu}, \bibinfo{person}{Stephanie Rogers}, \bibinfo{person}{Raymond Shiau}, \bibinfo{person}{Dmitry Kislyuk}, \bibinfo{person}{Kevin~C Ma}, \bibinfo{person}{Zhigang Zhong}, \bibinfo{person}{Jenny Liu}, {and} \bibinfo{person}{Yushi Jing}.} \bibinfo{year}{2017}\natexlab{}.
\newblock \showarticletitle{Related pins at pinterest: The evolution of a real-world recommender system}. In \bibinfo{booktitle}{\emph{Proceedings of the 26th international conference on world wide web companion}}. \bibinfo{pages}{583--592}.
\newblock


\bibitem[Liu et~al\mbox{.}(2018)]%
        {liu2018fast}
\bibfield{author}{\bibinfo{person}{Hongfu Liu}, \bibinfo{person}{Ziming Huang}, \bibinfo{person}{Qi Chen}, \bibinfo{person}{Mingqin Li}, \bibinfo{person}{Yun Fu}, {and} \bibinfo{person}{Lintao Zhang}.} \bibinfo{year}{2018}\natexlab{}.
\newblock \showarticletitle{Fast clustering with flexible balance constraints}. In \bibinfo{booktitle}{\emph{2018 IEEE International Conference on Big Data (Big Data)}}. IEEE, \bibinfo{pages}{743--750}.
\newblock


\bibitem[Liu et~al\mbox{.}(2022)]%
        {liu2022monolith}
\bibfield{author}{\bibinfo{person}{Zhuoran Liu}, \bibinfo{person}{Leqi Zou}, \bibinfo{person}{Xuan Zou}, \bibinfo{person}{Caihua Wang}, \bibinfo{person}{Biao Zhang}, \bibinfo{person}{Da Tang}, \bibinfo{person}{Bolin Zhu}, \bibinfo{person}{Yijie Zhu}, \bibinfo{person}{Peng Wu}, \bibinfo{person}{Ke Wang}, {and} \bibinfo{person}{Youlong Cheng}.} \bibinfo{year}{2022}\natexlab{}.
\newblock \showarticletitle{Monolith: Real Time Recommendation System With Collisionless Embedding Table}. In \bibinfo{booktitle}{\emph{5th Workshop on Online Recommender Systems and User Modeling (ORSUM2022), in conjunction with the 16th ACM Conference on Recommender Systems}}.
\newblock


\bibitem[Malkov and Yashunin(2018)]%
        {noauthor_hnswlib}
\bibfield{author}{\bibinfo{person}{Yu~A Malkov} {and} \bibinfo{person}{Dmitry~A Yashunin}.} \bibinfo{year}{2018}\natexlab{}.
\newblock \showarticletitle{Efficient and robust approximate nearest neighbor search using hierarchical navigable small world graphs}.
\newblock \bibinfo{journal}{\emph{IEEE transactions on pattern analysis and machine intelligence}} \bibinfo{volume}{42}, \bibinfo{number}{4} (\bibinfo{year}{2018}), \bibinfo{pages}{824--836}.
\newblock


\bibitem[Minaee et~al\mbox{.}(2021)]%
        {minaee2021deep}
\bibfield{author}{\bibinfo{person}{Shervin Minaee}, \bibinfo{person}{Nal Kalchbrenner}, \bibinfo{person}{Erik Cambria}, \bibinfo{person}{Narjes Nikzad}, \bibinfo{person}{Meysam Chenaghlu}, {and} \bibinfo{person}{Jianfeng Gao}.} \bibinfo{year}{2021}\natexlab{}.
\newblock \showarticletitle{Deep learning--based text classification: a comprehensive review}.
\newblock \bibinfo{journal}{\emph{ACM Computing Surveys (CSUR)}} \bibinfo{volume}{54}, \bibinfo{number}{3} (\bibinfo{year}{2021}), \bibinfo{pages}{1--40}.
\newblock


\bibitem[Mohoney et~al\mbox{.}(2024)]%
        {mohoney2024incremental}
\bibfield{author}{\bibinfo{person}{Jason Mohoney}, \bibinfo{person}{Anil Pacaci}, \bibinfo{person}{Shihabur~Rahman Chowdhury}, \bibinfo{person}{Umar~Farooq Minhas}, \bibinfo{person}{Jeffery Pound}, \bibinfo{person}{Cedric Renggli}, \bibinfo{person}{Nima Reyhani}, \bibinfo{person}{Ihab~F. Ilyas}, \bibinfo{person}{Theodoros Rekatsinas}, {and} \bibinfo{person}{Shivaram Venkataraman}.} \bibinfo{year}{2024}\natexlab{}.
\newblock \bibinfo{title}{Incremental IVF Index Maintenance for Streaming Vector Search}.
\newblock
\newblock
\showeprint[arxiv]{2411.00970}~[cs.DB]
\urldef\tempurl%
\url{https://arxiv.org/abs/2411.00970}
\showURL{%
\tempurl}


\bibitem[Mohoney et~al\mbox{.}(2023)]%
        {mohoney_high-throughput_2023}
\bibfield{author}{\bibinfo{person}{Jason Mohoney}, \bibinfo{person}{Anil Pacaci}, \bibinfo{person}{Shihabur~Rahman Chowdhury}, \bibinfo{person}{Ali Mousavi}, \bibinfo{person}{Ihab~F. Ilyas}, \bibinfo{person}{Umar~Farooq Minhas}, \bibinfo{person}{Jeffrey Pound}, {and} \bibinfo{person}{Theodoros Rekatsinas}.} \bibinfo{year}{2023}\natexlab{}.
\newblock \showarticletitle{High-{Throughput} {Vector} {Similarity} {Search} in {Knowledge} {Graphs}}.
\newblock \bibinfo{journal}{\emph{Proceedings of the ACM on Management of Data}} \bibinfo{volume}{1}, \bibinfo{number}{2} (\bibinfo{date}{June} \bibinfo{year}{2023}), \bibinfo{pages}{1--25}.
\newblock
\showISSN{2836-6573}
\urldef\tempurl%
\url{https://doi.org/10.1145/3589777}
\showDOI{\tempurl}


\bibitem[Okura et~al\mbox{.}(2017)]%
        {okura2017embedding}
\bibfield{author}{\bibinfo{person}{Shumpei Okura}, \bibinfo{person}{Yukihiro Tagami}, \bibinfo{person}{Shingo Ono}, {and} \bibinfo{person}{Akira Tajima}.} \bibinfo{year}{2017}\natexlab{}.
\newblock \showarticletitle{Embedding-based news recommendation for millions of users}. In \bibinfo{booktitle}{\emph{Proceedings of the 23rd ACM SIGKDD international conference on knowledge discovery and data mining}}. \bibinfo{pages}{1933--1942}.
\newblock


\bibitem[Pal et~al\mbox{.}(2020)]%
        {pal2020pinnersage}
\bibfield{author}{\bibinfo{person}{Aditya Pal}, \bibinfo{person}{Chantat Eksombatchai}, \bibinfo{person}{Yitong Zhou}, \bibinfo{person}{Bo Zhao}, \bibinfo{person}{Charles Rosenberg}, {and} \bibinfo{person}{Jure Leskovec}.} \bibinfo{year}{2020}\natexlab{}.
\newblock \showarticletitle{Pinnersage: Multi-modal user embedding framework for recommendations at pinterest}. In \bibinfo{booktitle}{\emph{Proceedings of the 26th ACM SIGKDD International Conference on Knowledge Discovery \& Data Mining}}. \bibinfo{pages}{2311--2320}.
\newblock


\bibitem[Pan et~al\mbox{.}(2024)]%
        {pan2024survey}
\bibfield{author}{\bibinfo{person}{James~Jie Pan}, \bibinfo{person}{Jianguo Wang}, {and} \bibinfo{person}{Guoliang Li}.} \bibinfo{year}{2024}\natexlab{}.
\newblock \showarticletitle{Survey of vector database management systems}.
\newblock \bibinfo{journal}{\emph{The VLDB Journal}} \bibinfo{volume}{33}, \bibinfo{number}{5} (\bibinfo{year}{2024}), \bibinfo{pages}{1591--1615}.
\newblock


\bibitem[Patel et~al\mbox{.}(2024)]%
        {patel2024acorn}
\bibfield{author}{\bibinfo{person}{Liana Patel}, \bibinfo{person}{Peter Kraft}, \bibinfo{person}{Carlos Guestrin}, {and} \bibinfo{person}{Matei Zaharia}.} \bibinfo{year}{2024}\natexlab{}.
\newblock \showarticletitle{ACORN: Performant and Predicate-Agnostic Search Over Vector Embeddings and Structured Data}.
\newblock \bibinfo{journal}{\emph{Proc. ACM Manag. Data}} \bibinfo{volume}{2}, \bibinfo{number}{3}, Article \bibinfo{articleno}{120} (\bibinfo{date}{May} \bibinfo{year}{2024}), \bibinfo{numpages}{27}~pages.
\newblock
\urldef\tempurl%
\url{https://doi.org/10.1145/3654923}
\showDOI{\tempurl}


\bibitem[Qin et~al\mbox{.}(2021)]%
        {qin2021mixer}
\bibfield{author}{\bibinfo{person}{An Qin}, \bibinfo{person}{Mengbai Xiao}, \bibinfo{person}{Yongwei Wu}, \bibinfo{person}{Xinjie Huang}, {and} \bibinfo{person}{Xiaodong Zhang}.} \bibinfo{year}{2021}\natexlab{}.
\newblock \showarticletitle{Mixer: efficiently understanding and retrieving visual content at web-scale}.
\newblock \bibinfo{journal}{\emph{Proceedings of the VLDB Endowment}} \bibinfo{volume}{14}, \bibinfo{number}{12} (\bibinfo{year}{2021}), \bibinfo{pages}{2906--2917}.
\newblock


\bibitem[Radford et~al\mbox{.}(2021a)]%
        {radford2021learning}
\bibfield{author}{\bibinfo{person}{Alec Radford}, \bibinfo{person}{Jong~Wook Kim}, \bibinfo{person}{Chris Hallacy}, \bibinfo{person}{Aditya Ramesh}, \bibinfo{person}{Gabriel Goh}, \bibinfo{person}{Sandhini Agarwal}, \bibinfo{person}{Girish Sastry}, \bibinfo{person}{Amanda Askell}, \bibinfo{person}{Pamela Mishkin}, \bibinfo{person}{Jack Clark}, {et~al\mbox{.}}} \bibinfo{year}{2021}\natexlab{a}.
\newblock \showarticletitle{Learning transferable visual models from natural language supervision}. In \bibinfo{booktitle}{\emph{International conference on machine learning}}. PMLR, \bibinfo{pages}{8748--8763}.
\newblock


\bibitem[Radford et~al\mbox{.}(2021b)]%
        {pmlr-v139-radford21a}
\bibfield{author}{\bibinfo{person}{Alec Radford}, \bibinfo{person}{Jong~Wook Kim}, \bibinfo{person}{Chris Hallacy}, \bibinfo{person}{Aditya Ramesh}, \bibinfo{person}{Gabriel Goh}, \bibinfo{person}{Sandhini Agarwal}, \bibinfo{person}{Girish Sastry}, \bibinfo{person}{Amanda Askell}, \bibinfo{person}{Pamela Mishkin}, \bibinfo{person}{Jack Clark}, \bibinfo{person}{Gretchen Krueger}, {and} \bibinfo{person}{Ilya Sutskever}.} \bibinfo{year}{2021}\natexlab{b}.
\newblock \showarticletitle{Learning Transferable Visual Models From Natural Language Supervision}. In \bibinfo{booktitle}{\emph{Proceedings of the 38th International Conference on Machine Learning}} \emph{(\bibinfo{series}{Proceedings of Machine Learning Research}, Vol.~\bibinfo{volume}{139})}, \bibfield{editor}{\bibinfo{person}{Marina Meila} {and} \bibinfo{person}{Tong Zhang}} (Eds.). \bibinfo{publisher}{PMLR}, \bibinfo{pages}{8748--8763}.
\newblock
\urldef\tempurl%
\url{https://proceedings.mlr.press/v139/radford21a.html}
\showURL{%
\tempurl}


\bibitem[Sculley(2010)]%
        {sculley2010web}
\bibfield{author}{\bibinfo{person}{David Sculley}.} \bibinfo{year}{2010}\natexlab{}.
\newblock \showarticletitle{Web-scale k-means clustering}. In \bibinfo{booktitle}{\emph{Proceedings of the 19th international conference on World wide web}}. \bibinfo{pages}{1177--1178}.
\newblock


\bibitem[Selinger et~al\mbox{.}(1979)]%
        {selinger1979access}
\bibfield{author}{\bibinfo{person}{P.~Griffiths Selinger}, \bibinfo{person}{M.~M. Astrahan}, \bibinfo{person}{D.~D. Chamberlin}, \bibinfo{person}{R.~A. Lorie}, {and} \bibinfo{person}{T.~G. Price}.} \bibinfo{year}{1979}\natexlab{}.
\newblock \showarticletitle{Access path selection in a relational database management system}. In \bibinfo{booktitle}{\emph{Proceedings of the 1979 ACM SIGMOD International Conference on Management of Data}} (Boston, Massachusetts) \emph{(\bibinfo{series}{SIGMOD '79})}. \bibinfo{publisher}{Association for Computing Machinery}, \bibinfo{address}{New York, NY, USA}, \bibinfo{pages}{23–34}.
\newblock
\showISBNx{089791001X}
\urldef\tempurl%
\url{https://doi.org/10.1145/582095.582099}
\showDOI{\tempurl}


\bibitem[Simhadri et~al\mbox{.}(2024)]%
        {simhadri2024results}
\bibfield{author}{\bibinfo{person}{Harsha~Vardhan Simhadri}, \bibinfo{person}{Martin Aum{\"u}ller}, \bibinfo{person}{Amir Ingber}, \bibinfo{person}{Matthijs Douze}, \bibinfo{person}{George Williams}, \bibinfo{person}{Magdalen~Dobson Manohar}, \bibinfo{person}{Dmitry Baranchuk}, \bibinfo{person}{Edo Liberty}, \bibinfo{person}{Frank Liu}, \bibinfo{person}{Ben Landrum}, {et~al\mbox{.}}} \bibinfo{year}{2024}\natexlab{}.
\newblock \showarticletitle{Results of the Big ANN: NeurIPS'23 competition}.
\newblock \bibinfo{journal}{\emph{arXiv preprint arXiv:2409.17424}} (\bibinfo{year}{2024}).
\newblock


\bibitem[Singh et~al\mbox{.}(2021)]%
        {singh2021freshdiskann}
\bibfield{author}{\bibinfo{person}{Aditi Singh}, \bibinfo{person}{Suhas~Jayaram Subramanya}, \bibinfo{person}{Ravishankar Krishnaswamy}, {and} \bibinfo{person}{Harsha~Vardhan Simhadri}.} \bibinfo{year}{2021}\natexlab{}.
\newblock \showarticletitle{FreshDiskANN: {A} Fast and Accurate Graph-Based {ANN} Index for Streaming Similarity Search}.
\newblock \bibinfo{journal}{\emph{CoRR}}  \bibinfo{volume}{abs/2105.09613} (\bibinfo{year}{2021}).
\newblock
\showeprint[arXiv]{2105.09613}
\urldef\tempurl%
\url{https://arxiv.org/abs/2105.09613}
\showURL{%
\tempurl}


\bibitem[Sundaram et~al\mbox{.}(2013)]%
        {sundaram_streaming_2013}
\bibfield{author}{\bibinfo{person}{Narayanan Sundaram}, \bibinfo{person}{Aizana Turmukhametova}, \bibinfo{person}{Nadathur Satish}, \bibinfo{person}{Todd Mostak}, \bibinfo{person}{Piotr Indyk}, \bibinfo{person}{Samuel Madden}, {and} \bibinfo{person}{Pradeep Dubey}.} \bibinfo{year}{2013}\natexlab{}.
\newblock \showarticletitle{Streaming similarity search over one billion tweets using parallel locality-sensitive hashing}.
\newblock \bibinfo{journal}{\emph{Proceedings of the VLDB Endowment}} \bibinfo{volume}{6}, \bibinfo{number}{14} (\bibinfo{date}{Sept.} \bibinfo{year}{2013}), \bibinfo{pages}{1930--1941}.
\newblock
\showISSN{2150-8097}
\urldef\tempurl%
\url{https://doi.org/10.14778/2556549.2556574}
\showDOI{\tempurl}


\bibitem[Wang et~al\mbox{.}(2018)]%
        {wang2018billion}
\bibfield{author}{\bibinfo{person}{Jizhe Wang}, \bibinfo{person}{Pipei Huang}, \bibinfo{person}{Huan Zhao}, \bibinfo{person}{Zhibo Zhang}, \bibinfo{person}{Binqiang Zhao}, {and} \bibinfo{person}{Dik~Lun Lee}.} \bibinfo{year}{2018}\natexlab{}.
\newblock \showarticletitle{Billion-scale Commodity Embedding for E-commerce Recommendation in Alibaba}. In \bibinfo{booktitle}{\emph{Proceedings of the 24th ACM SIGKDD International Conference on Knowledge Discovery \& Data Mining}} (London, United Kingdom) \emph{(\bibinfo{series}{KDD '18})}. \bibinfo{publisher}{Association for Computing Machinery}, \bibinfo{address}{New York, NY, USA}, \bibinfo{pages}{839–848}.
\newblock
\showISBNx{9781450355520}
\urldef\tempurl%
\url{https://doi.org/10.1145/3219819.3219869}
\showDOI{\tempurl}


\bibitem[Wang et~al\mbox{.}(2021)]%
        {wang2021milvus}
\bibfield{author}{\bibinfo{person}{Jianguo Wang}, \bibinfo{person}{Xiaomeng Yi}, \bibinfo{person}{Rentong Guo}, \bibinfo{person}{Hai Jin}, \bibinfo{person}{Peng Xu}, \bibinfo{person}{Shengjun Li}, \bibinfo{person}{Xiangyu Wang}, \bibinfo{person}{Xiangzhou Guo}, \bibinfo{person}{Chengming Li}, \bibinfo{person}{Xiaohai Xu}, \bibinfo{person}{Kun Yu}, \bibinfo{person}{Yuxing Yuan}, \bibinfo{person}{Yinghao Zou}, \bibinfo{person}{Jiquan Long}, \bibinfo{person}{Yudong Cai}, \bibinfo{person}{Zhenxiang Li}, \bibinfo{person}{Zhifeng Zhang}, \bibinfo{person}{Yihua Mo}, \bibinfo{person}{Jun Gu}, \bibinfo{person}{Ruiyi Jiang}, \bibinfo{person}{Yi Wei}, {and} \bibinfo{person}{Charles Xie}.} \bibinfo{year}{2021}\natexlab{}.
\newblock \showarticletitle{Milvus: A Purpose-Built Vector Data Management System}. In \bibinfo{booktitle}{\emph{Proceedings of the 2021 International Conference on Management of Data}} (Virtual Event, China) \emph{(\bibinfo{series}{SIGMOD '21})}. \bibinfo{publisher}{Association for Computing Machinery}, \bibinfo{address}{New York, NY, USA}, \bibinfo{pages}{2614–2627}.
\newblock
\showISBNx{9781450383431}
\urldef\tempurl%
\url{https://doi.org/10.1145/3448016.3457550}
\showDOI{\tempurl}


\bibitem[Wei et~al\mbox{.}(2020)]%
        {wei2020analyticdb}
\bibfield{author}{\bibinfo{person}{Chuangxian Wei}, \bibinfo{person}{Bin Wu}, \bibinfo{person}{Sheng Wang}, \bibinfo{person}{Renjie Lou}, \bibinfo{person}{Chaoqun Zhan}, \bibinfo{person}{Feifei Li}, {and} \bibinfo{person}{Yuanzhe Cai}.} \bibinfo{year}{2020}\natexlab{}.
\newblock \showarticletitle{Analyticdb-v: A hybrid analytical engine towards query fusion for structured and unstructured data}.
\newblock \bibinfo{journal}{\emph{Proceedings of the VLDB Endowment}} \bibinfo{volume}{13}, \bibinfo{number}{12} (\bibinfo{year}{2020}), \bibinfo{pages}{3152--3165}.
\newblock


\bibitem[Xu et~al\mbox{.}(2023)]%
        {xu_spfresh_2023}
\bibfield{author}{\bibinfo{person}{Yuming Xu}, \bibinfo{person}{Hengyu Liang}, \bibinfo{person}{Jin Li}, \bibinfo{person}{Shuotao Xu}, \bibinfo{person}{Qi Chen}, \bibinfo{person}{Qianxi Zhang}, \bibinfo{person}{Cheng Li}, \bibinfo{person}{Ziyue Yang}, \bibinfo{person}{Fan Yang}, \bibinfo{person}{Yuqing Yang}, \bibinfo{person}{Peng Cheng}, {and} \bibinfo{person}{Mao Yang}.} \bibinfo{year}{2023}\natexlab{}.
\newblock \showarticletitle{{SPFresh}: {Incremental} {In}-{Place} {Update} for {Billion}-{Scale} {Vector} {Search}}. In \bibinfo{booktitle}{\emph{Proceedings of the 29th {Symposium} on {Operating} {Systems} {Principles}}} \emph{(\bibinfo{series}{{SOSP} '23})}. \bibinfo{publisher}{Association for Computing Machinery}, \bibinfo{address}{New York, NY, USA}, \bibinfo{pages}{545--561}.
\newblock
\showISBNx{9798400702297}
\urldef\tempurl%
\url{https://doi.org/10.1145/3600006.3613166}
\showDOI{\tempurl}


\bibitem[Zheng et~al\mbox{.}(2020)]%
        {zheng2020pm}
\bibfield{author}{\bibinfo{person}{Bolong Zheng}, \bibinfo{person}{Zhao Xi}, \bibinfo{person}{Lianggui Weng}, \bibinfo{person}{Nguyen Quoc~Viet Hung}, \bibinfo{person}{Hang Liu}, {and} \bibinfo{person}{Christian~S Jensen}.} \bibinfo{year}{2020}\natexlab{}.
\newblock \showarticletitle{PM-LSH: A fast and accurate LSH framework for high-dimensional approximate NN search}.
\newblock \bibinfo{journal}{\emph{Proceedings of the VLDB Endowment}} \bibinfo{volume}{13}, \bibinfo{number}{5} (\bibinfo{year}{2020}), \bibinfo{pages}{643--655}.
\newblock


\end{thebibliography}
